\documentclass[iop]{emulateapj}

\newcommand{\ftwo}{$F_{E>200{\rm MeV}}$} 
\newcommand{\fone}{$F_{E>100{\rm MeV}}$} 
 
\newcommand{\phcms}{${\rm photons}\ {\rm cm}^{-2}\ {\rm s}^{-1}$} 
\newcommand{\Fermi}{{\it Fermi}}
\newcommand{\RXTE}{{\it RXTE}}
\newcommand{\Swift}{{\it Swift}}
\newcommand{\Suzaku}{{\it Suzaku}}
\newcommand{\Spitzer}{{\it Spitzer}}
\newcommand{\XMM}{XMM-\textit{Newton}}
\newcommand{\degree}{$^{\circ}$}

\shorttitle{Radio to high-energy $\gamma$-ray observations of 3C~279 in 2008-2010}
\shortauthors{Hayashida et al.}

\begin{document}

\pagenumbering{arabic}


\title{The structure and emission model of the relativistic jet in 
the quasar 3C~279 inferred from radio to high-energy $\gamma$-ray 
observations in 2008--2010}
\author{
M.~Hayashida\altaffilmark{1,2,3}, 
G.~M.~Madejski\altaffilmark{1,13}, 
K.~Nalewajko\altaffilmark{47,52, 53}, 
M.~Sikora\altaffilmark{52,54}, 
A.~E.~Wehrle\altaffilmark{20}, 
P.~Ogle\altaffilmark{55}, 
W.~Collmar\altaffilmark{7}, 
S.~Larsson\altaffilmark{10,11,12}, 
Y.~Fukazawa\altaffilmark{8}, 
R.~Itoh\altaffilmark{8},
J.~Chiang\altaffilmark{1}, 
{\L}.~Stawarz\altaffilmark{16,17}, 
R.~D.~Blandford\altaffilmark{1},
J.~L.~Richards\altaffilmark{15}, 
W.~Max-Moerbeck\altaffilmark{15}, 
A.~Readhead\altaffilmark{15}, 
R.~Buehler\altaffilmark{1}, 
E.~Cavazzuti\altaffilmark{4}, 
S.~Ciprini\altaffilmark{5,6}, 
N.~Gehrels\altaffilmark{9}, 
A.~Reimer\altaffilmark{14,1}, 
A.~Szostek\altaffilmark{1,17}, 
T.~Tanaka\altaffilmark{1}, 
G.~Tosti\altaffilmark{18,6}, 
Y.~Uchiyama\altaffilmark{1},
K.~S.~Kawabata\altaffilmark{40}, 
M.~Kino\altaffilmark{24}, 
K.~Sakimoto\altaffilmark{8}, 
M.~Sasada\altaffilmark{8}, 
S.~Sato\altaffilmark{24}, 
M.~Uemura\altaffilmark{40},
M.~Yamanaka\altaffilmark{8} 
J.~Greiner\altaffilmark{7}, 
T.~Kruehler\altaffilmark{42}, 
A.~Rossi\altaffilmark{49}, 
J.~P.~Macquart\altaffilmark{45}, 
D.~C.-J.~Bock\altaffilmark{28},
M.~Villata\altaffilmark{19}, 
C.~M.~Raiteri\altaffilmark{19}, 
I.~Agudo\altaffilmark{21, 27}, 
H.~D.~Aller\altaffilmark{22}, 
M.~F.~Aller\altaffilmark{22}, 
A.~A.~Arkharov\altaffilmark{23}, 
U.~Bach\altaffilmark{51}, 
E.~Ben\'itez\altaffilmark{25}, 
A.~Berdyugin\altaffilmark{26}, 
D.~A.~Blinov\altaffilmark{35}, 
K.~Blumenthal\altaffilmark{27}, 
M.~B\"ottcher\altaffilmark{29}, 
C.~S.~Buemi\altaffilmark{30}, 
D.~Carosati\altaffilmark{31, 56}, 
W.~P.~Chen\altaffilmark{32}, 
A.~Di~Paola\altaffilmark{33}, 
M.~Dolci\altaffilmark{34}, 
N.~V.~Efimova\altaffilmark{23,35}, 
E.~Forn\'e\altaffilmark{57},
J.~L.~G\'omez\altaffilmark{21}, 
M.~A.~Gurwell\altaffilmark{36}, 
J.~Heidt\altaffilmark{37}, 
D.~Hiriart\altaffilmark{38}, 
B.~Jordan\altaffilmark{39}, 
S.~G.~Jorstad\altaffilmark{27, 35}, 
M.~Joshi\altaffilmark{27}, 
G.~Kimeridze\altaffilmark{41}, 
T.~S.~Konstantinova\altaffilmark{35}, 
E.~N.~Kopatskaya\altaffilmark{35}, 
E.~Koptelova\altaffilmark{32, 58}, 
O.~M.~Kurtanidze\altaffilmark{41}, 
A.~L\"ahteenm\"aki\altaffilmark{43}, 
A.~Lamerato\altaffilmark{29}, 
V.~M.~Larionov\altaffilmark{35, 23, 44}, 
E.~G.~Larionova\altaffilmark{35}, 
L.~V.~Larionova\altaffilmark{35}, 
P.~Leto\altaffilmark{30}, 
E.~Lindfors\altaffilmark{26}, 
A.~P.~Marscher\altaffilmark{27}, 
I.~M.~McHardy\altaffilmark{46}, 
S.~N.~Molina\altaffilmark{21}, 
D.~A.~Morozova\altaffilmark{35}, 
M.~G.~Nikolashvili\altaffilmark{41}, 
K.~Nilsson\altaffilmark{48}, 
R.~Reinthal\altaffilmark{26}, 
P.~Roustazadeh\altaffilmark{29}, 
T.~Sakamoto\altaffilmark{9}, 
L.~A.~Sigua\altaffilmark{41}, 
A.~Sillanp\"a\"a\altaffilmark{26}, 
L.~Takalo\altaffilmark{26}, 
J.~Tammi\altaffilmark{43}, 
B.~Taylor\altaffilmark{27, 50}, 
M.~Tornikoski\altaffilmark{43}, 
C.~Trigilio\altaffilmark{30}, 
I.~S.~Troitsky\altaffilmark{35}, 
G.~Umana\altaffilmark{30}
}
\altaffiltext{1}{Kavli Institute for Particle Astrophysics and Cosmology, SLAC National Accelerator Laboratory,  Stanford University,
2575 Sand Hill Road M/S 29, Menlo Park, CA 94025, USA}
\altaffiltext{2}{Department of Astronomy, Graduate School of Science, Kyoto University, Sakyo-ku, Kyoto 606-8502, Japan}
\altaffiltext{3}{email: mahaya@slac.stanford.edu}
\altaffiltext{4}{Agenzia Spaziale Italiana (ASI) Science Data Center, I-00044 Frascati (Roma), Italy}
\altaffiltext{5}{ASI Science Data Center, I-00044 Frascati (Roma), Italy}
\altaffiltext{6}{Dipartimento di Fisica, Universit\`a degli Studi di Perugia, I-06123 Perugia, Italy}
\altaffiltext{7}{Max-Planck Institut f\"ur extraterrestrische Physik, 85748 Garching, Germany}
\altaffiltext{8}{Department of Physical Sciences, Hiroshima University, Higashi-Hiroshima, Hiroshima 739-8526, Japan}
\altaffiltext{9}{NASA Goddard Space Flight Center, Greenbelt, MD 20771, USA}
\altaffiltext{10}{Department of Physics, Stockholm University, AlbaNova, SE-106 91 Stockholm, Sweden}
\altaffiltext{11}{The Oskar Klein Centre for Cosmoparticle Physics, AlbaNova, SE-106 91 Stockholm, Sweden}
\altaffiltext{12}{Department of Astronomy, Stockholm University, SE-106 91 Stockholm, Sweden}
\altaffiltext{13}{email: madejski@slac.stanford.edu}
\altaffiltext{14}{Institut f\"ur Astro- und Teilchenphysik and Institut f\"ur Theoretische Physik, Leopold-Franzens-Universit\"at Innsbruck, A-6020 Innsbruck, Austria}
\altaffiltext{15}{Cahill Center for Astronomy and Astrophysics, California Institute of Technology, Pasadena, CA 91125, USA}
\altaffiltext{16}{Institute of Space and Astronautical Science, JAXA, 3-1-1 Yoshinodai, Chuo-ku, Sagamihara, Kanagawa 252-5210, Japan}
\altaffiltext{17}{Astronomical Observatory, Jagiellonian University, 30-244 Krak\'ow, Poland}
\altaffiltext{18}{Istituto Nazionale di Fisica Nucleare, Sezione di Perugia, I-06123 Perugia, Italy}
\altaffiltext{19}{INAF, Osservatorio Astronomico di Torino, I-10025 Pino Torinese (TO), Italy}
\altaffiltext{20}{Space Science Institute, Boulder, CO 80301, USA}
\altaffiltext{21}{Instituto de Astrof\'isica de Andaluc\'ia, CSIC, E-18080 Granada, Spain}
\altaffiltext{22}{Department of Astronomy, University of Michigan, Ann Arbor, MI 48109-1090, USA}
\altaffiltext{23}{Pulkovo Observatory, 196140 St. Petersburg, Russia}
\altaffiltext{24}{Department of Physics and Astrophysics, Nagoya University, Chikusa-ku Nagoya 464-8602, Japan}
\altaffiltext{25}{Instituto de Astronom\'ia, Universidad Nacional Aut\'onoma de M\'exico, M\'exico, D. F., M\'exico}
\altaffiltext{26}{Tuorla Observatory, Department of Physics and Astronomy, University of Turku, FI-21500 Piikki\"o, Finland}
\altaffiltext{27}{Institute for Astrophysical Research, Boston University, Boston, MA 02215, USA}
\altaffiltext{28}{CSIRO Astronomy and Space Science P.O. Box 76, Epping, NSW 1710, Australia}
\altaffiltext{29}{Astrophysical Institute Department of Physics and Astronomy, Ohio University, Athens, OH 45701, USA}
\altaffiltext{30}{INAF, Osservatorio Astrofisico di Catania, 95123 Catania, Italy}
\altaffiltext{31}{EPT Observatories, Tijarafe, La Palma, Spain}
\altaffiltext{32}{Graduate Institute of Astronomy, National Central University, Jhongli 32001, Taiwan}
\altaffiltext{33}{INAF, Osservatorio Astronomico di Roma, I-00040 Monte Porzio Catone (Roma), Italy}
\altaffiltext{34}{INAF, Osservatorio Astronomico di Collurania ``Vincenzo Cerruli", 64100 Teramo, Italy}
\altaffiltext{35}{Astronomical Institute, St. Petersburg State University, St. Petersburg, Russia}
\altaffiltext{36}{Harvard-Smithsonian Center for Astrophysics, Cambridge, MA 02138, USA}
\altaffiltext{37}{ZAH, Landessternwarte, Universit\"at Heidelberg, K\"onigstuhl, D 69117 Heidelberg, Germany}
\altaffiltext{38}{Instituto de Astronom\'ia, Universidad Nacional Aut\'onoma de M\'exico, Ensenada, B. C., M\'exico}
\altaffiltext{39}{School of Cosmic Physics, Dublin Institute for Advanced Studies, Dublin, 2, Ireland}
\altaffiltext{40}{Hiroshima Astrophysical Science Center, Hiroshima University, Higashi-Hiroshima, Hiroshima 739-8526, Japan}
\altaffiltext{41}{Abastumani Observatory, Mt. Kanobili, 0301 Abastumani, Georgia}
\altaffiltext{42}{Dark Cosmology Centre, Niels Bohr Institute, University of Copenhagen, 2100 Copenhagen, Denmark}
\altaffiltext{43}{Aalto University Mets\"ahovi Radio Observatory, FIN-02540 Kylm\"al\"a, Finland}
\altaffiltext{44}{Isaac Newton Institute of Chile, St. Petersburg Branch, St. Petersburg, Russia}
\altaffiltext{45}{International Centre for Radio Astronomy Research and Curtin University of Technology, Bentley, WA 6845, Australia}
\altaffiltext{46}{School of Physics and Astronomy, University of Southampton, Highfield, Southampton, SO17 1BJ, UK}
\altaffiltext{47}{University of Colorado, UCB 440, Boulder, CO 80309, USA}
\altaffiltext{48}{Finnish Centre for Astronomy with ESO (FINCA), University of Turku, FI-21500 Piikii\"o, Finland}
\altaffiltext{49}{Thuringer Landessternwarte Tautenburg, D-07778 Tautenburg, Germany}
\altaffiltext{50}{Lowell Observatory, Flagstaff, AZ 86001, USA}
\altaffiltext{51}{Max-Planck-Institut f\"ur Radioastronomie, Auf dem H\"ugel 69, 53121 Bonn, Germany}
\altaffiltext{52}{Nicolaus Copernicus Astronomical Center, 00-716 Warsaw, Poland}
\altaffiltext{53}{email: knalew@Colorado.edu}
\altaffiltext{54}{email: sikora@camk.edu.pl}
\altaffiltext{55}{Infrared Processing and Analysis Center, California Institute of Technology Pasadena, CA 91125}
\altaffiltext{56}{INAF, TNG Fundacio\'n Galileo Galilei, La Palma, Spain}
\altaffiltext{57}{Agrupaci\'o Astron\`omica de Sabadell, 08206 Sabadell, Spain}
\altaffiltext{58}{Department of Physics, National Taiwan University, 106 Taipei, Taiwan}


%


\begin{abstract}
We present time-resolved broad-band observations of the quasar 3C~279 
obtained from multi-wavelength campaigns conducted during the first 
two years of the \Fermi\ Gamma-ray Space Telescope mission.  
While investigating the previously reported $\gamma$-ray/optical 
flare accompanied by a change in optical polarization, 
we found that the optical emission appears delayed with respect 
to the $\gamma$-ray emission by about 10 days.  X-ray observations 
reveal a pair of `isolated' flares separated by $\sim 90$ days, 
with only weak $\gamma$-ray/optical counterparts.
The spectral structure measured by \Spitzer\ 
reveals a synchrotron component peaking in the mid-infrared band 
with a sharp break at the far-infrared band during the $\gamma$-ray flare,
while the peak appears in the mm/sub-mm band in the low state.
Selected spectral energy distributions are fitted with 
leptonic models including Comptonization 
of external radiation produced in a dusty torus 
or the broad-line region.
Adopting the interpretation of the polarization swing involving 
propagation of the emitting region along a curved trajectory, 
we can explain the evolution of the broad-band spectra 
during the $\gamma$-ray flaring event by a shift of its location from
$\sim 1$\,pc to $\sim 4$\,pc from the central black hole.
On the other hand, if the $\gamma$-ray flare is generated instead 
at sub-pc distance from the central black hole, 
the far-infrared break can be explained by synchrotron self-absorption.
We also model the low spectral state, dominated by the mm/sub-mm peaking synchrotron component, 
and suggest that the corresponding inverse-Compton component explains the steady X-ray emission.

\end{abstract}



\section{Introduction}
Blazars are Active Galactic Nuclei (AGN) characterized by highly luminous 
and rapidly variable continuum emission at all observed bands. The most commonly 
accepted scenario has their broad-band emission Doppler-boosted by
a relativistic jet pointing close to our line of sight~\citep[e.g.,][]{Ulr97}.  While the jet emission 
usually dominates the observed broad-band spectrum, 
the optical/ultra-violet (UV) and infrared (IR) spectra 
often also reveal signatures of the central engine:  broad emission lines, and 
in some cases, quasi-thermal optical/UV emission and IR dust emission,
indicating the presence of 
an accreting supermassive black hole. Most viable current models for the origin of 
such jets involve conversion of the gravitational energy of matter flowing onto 
the black hole to the kinetic energy of the relativistic outflow or tapping
the rotation energy of a spinning black hole. However, the 
conversion process itself is not well understood, and many additional questions 
regarding the dissipation region of the jet's energy into radiation 
and, in particular, its location remain unanswered. 

Major advances in understanding of blazars came as a result of the discovery 
by the EGRET instrument on board {\it Compton Gamma-Ray Observatory 
(CGRO)} that they are strong $\gamma$-ray emitters, with $\gamma$-rays dominating 
radiative output~\citep{3EG}. With this, 
multi-band observations including the $\gamma$-ray band
hold the promise of answering many outstanding
questions regarding the structure of the relativistic jets of blazars.  

3C~279~\citep[$z=0.536$;][]{Lyn65} is in fact one of the first 
$\gamma$-ray blazars discovered by EGRET in 1991~\citep{Har92}.  The 
$\gamma$-ray signal had been significantly detected in each observation by 
EGRET since its discovery~\citep[see, e.g.,][]{Har01SED}, 
with the flux having ranged over roughly 2 orders of 
magnitude, from $\sim 10^{-7}$ up to $\sim 10^{-5}$ {\phcms} 
above 100 MeV~\citep{Mar94, Weh98}, and a factor of 2 variation 
on timescales as short as 8 hrs.  The photon index in the EGRET $\gamma$-ray 
band ranged from 1.8 to 2.3~\citep{Nan07}.  On a few occasions, 3C~279 was 
also detected at lower energies by {\it CGRO's} OSSE (50~keV$-$1 MeV)~\citep{OSSE} 
and COMPTEL (0.75$-$30 MeV)~\citep{Her93, Col01} instruments, indicating 
that the $\gamma$-ray emission forms a broad peak in the $\nu F_{\nu}$ 
representation.
In 2008 July, the AGILE satellite observed a $\gamma$-ray flare associated with the source 
with $11.1\,\sigma$ significance~\citep{AGILE}, with an average flux above 
100\,MeV of $(21.0\pm3.8)\times10^{-7}$ \phcms~and the photon index of 
$2.22\pm0.23$ between 100 and 1000\,MeV.
In the very-high-energy (VHE) $\gamma$-ray regime above 100~GeV, the imaging atmospheric 
Cherenkov telescope MAGIC detected flares twice in 2006 February~\citep{MAGIC}
and in 2007 January~\citep{MAGICUL}, which made this source the most distant currently known VHE $\gamma$-ray emitter.  

Optical and UV observations of the source in relatively low states - when the 
jet emission was relatively faint - allowed a study of the accreting black hole
and the associated accretion disk. The luminosity of the accretion disk was 
estimated to be $L_{D} \sim 2\times10^{45}$ erg s$^{-1}$~\citep{Pia99}.
The mass of central supermassive black hole was estimated to 
be in the range of (3--8)$\times 10^{8} M_{\Sun}$
using the luminosity of optical broad line~\citep{Woo02} or
the $H_{\beta}$ 
line width~\citep{Gu01}. Those values are similar to 
the estimates based on the luminosity of the host galaxy~\citep{Nil09}.  

3C~279 contains a compact radio core, associated with time-variable 
jet-like structure.  Radio observations at 43 GHz by Very Long 
Baseline Array (VLBA) between 1998 March and 2001 April revealed 
superluminal motion of the jet with apparent speeds that range from 
$5c$ to $17c$~\citep{Jor04, Jor05}.  Those observations 
also allowed an estimate of the Lorentz factor of the jet flow of $\Gamma_{\rm j} = 15.5\pm2.5$ 
and of the viewing angle of the jet $\Theta_0 = 2.1\pm1.1$ deg, 
corresponding to a Doppler beaming factor of $\delta = 24.1\pm6.5$.
A change of the trajectory of a jet component has also been reported in radio observations with VLBA~\citep{Hom03}.
Those authors estimated the jet component to be moving with a Lorentz factor $\Gamma_{\rm j} \gtrsim 15$ at
an initial viewing angle of $\lesssim 1$\degree.

The broad-band spectral energy distribution (SED) of the source 
is characterized by a 
two-bump structure, similar to many other $\gamma$-ray blazars.  
In the context of widely accepted leptonic models, the lower-frequency bump, 
peaking at the far-IR and extending to the extreme UV band, is commonly
 ascribed to synchrotron radiation from relativistic electrons 
in the jet. The second bump, spanning from the X-ray to the 
$\gamma$-ray band with a peak in the MeV-GeV range, is believed to 
be generated via inverse-Compton scattering, presumably by the same 
population of particles that radiate at lower energies via the 
synchrotron process.  The seed photons for the Compton scattering 
can be synchrotron photons~\citep[synchrotron self-Compton: SSC,][]{Mar92, Blo96}, 
accretion disk photons~\citep[external 
Compton scattering of direct disk radiation: ECD,][]{Der92, 
Der93} and accretion disk photons re-scattered by the broad-line 
region clouds/intercloud medium~\citep[ECC,][]{Sik94, Bla95}, or infrared 
radiation from a torus located beyond the broad-line region~\citep[ERC-IR,][]{Sik94}.  
Specific to 3C~279, multiwavelength snapshot observations for 
several epochs including $\gamma$-rays were presented in~\citet{Har01SED}. 
Those authors explained the overall spectra using the leptonic model, where 
the X-ray photons are mainly produced by SSC, and both ECD and ECC 
contribute to the $\gamma$-ray emission. 
Spectral variability was explained by 
variations of the 
bulk Lorentz factor of the jet, accompanied by changes in 
the spectral shape of the electron distribution.  

The optical variability of 3C~279 is extreme: in 1937, it showed optical ($B$) 
magnitude of 11.27~\citep{Eac75}, making 
it one of the most luminous active galaxies ever recorded.
The strong variability recorded 
in all bands provides an opportunity to 
establish the relationship between emission in those bands, 
and thus 
can be used to constrain theoretical models of 
physical regions of the jet responsible for such emission.  Many such 
multi-wavelength campaigns have been conducted~\citep[see, 
e.g.,][]{Mar94, Weh98, Lar08, Col10} but they have not revealed 
a simple relationship between the variability in various bands:  radiation in 
different spectral regimes does not always rise and fall simultaneously, although the 
periods of increased rapid activity in all bands seem to last for several months, and 
take place when the source is relatively bright.  A recent paper by \citet{Cha08} presents 
the results of the monitoring of 3C~279 for 11 years in radio, optical and 
X-rays, and discusses the details of the jet structure based on multi-band 
correlation studies. However, due to the lack of deployed instruments, long term 
monitoring observations could not include the $\gamma$-ray regime, where  
the source often shows stronger variability than in other bands.  

The launch of the {\it Fermi} Gamma-ray Space Telescope on 2008 June 11 
has rejuvenated multi-band studies of blazars. 
The Large Area Telescope~\citep[LAT;][]{LAT} instrument on {\it Fermi} 
can monitor all $\gamma$-ray sources on the sky with its
a wide filed of view and a much larger effective area compared to earlier $\gamma$-ray missions. 
Taking advantage of this new instrument for $\gamma$-ray observations,
we have organized intensive multiwavelength campaigns for 
3C~279 from radio to the high-energy $\gamma$-ray energy ranges.  Many ground-based telescopes 
(cm, mm, near-IR and optical) and various satellites 
(IR, UV, X-ray, hard X-ray and high-energy $\gamma$-ray)
participated in this campaign. We 
reported the first results of the campaign in~\citet[][hereafter Paper I]{paper1}, 
where we discovered the dramatic change of the optical polarization 
coincident with the $\gamma$-ray flare. Here, 
we provide details of the multi-band observations and the 
interpretation of those data for the 2-year interval 
between 2008 August and 2010 August. 
In Section \ref{section:fermi}, we present and briefly discuss the features 
of the LAT $\gamma$-ray data;  in Section \ref{section:mwl}, we present the data in 
lower energy bands.  Section \ref{section:results} highlights the features of time series 
measured in various bands including their cross correlations, 
and the general properties of the broad-band spectral energy distribution.  
In Section \ref{section:modeling}, we provide viable emission models for the source in the context of leptonic scenarios.

\section{\Fermi-LAT data and results}
\label{section:fermi}

\Fermi-LAT is a pair-production telescope with large effective area (8,000 cm$^2$ on axis at 1\,GeV for the event class considered here),
and large field of view (2.4\,sr at 1\,GeV), sensitive to $\gamma$ rays in the energy range from $20$\,MeV to $> 300$\,GeV. 
Information regarding on-orbit calibration procedures is given in \citet{Calib}.
\Fermi-LAT normally operates in a scanning `sky-survey' mode, which provides a full-sky coverage every two orbits (3 hours). For operational reasons, the standard rocking angle (defined as the angle between the zenith and the center of the LAT field of view) for survey mode was increased from 35\degree\ to 50\degree\ on 2009 September 3.

\subsection{Observation and data reductions}

The data used here comprise 2-year observations obtained between 
2008 August 4 and 2010 August 6 (MJD 54682 - 55414).
We used the standard LAT analysis software, \textit{ScienceTools v9r21}. 
The events were selected using
so-called ``diffuse class" events. In addition, 
we excluded the events with zenith angles greater than $100^{\circ}$ to 
avoid the contamination of 
the Earth-limb secondary $\gamma$ radiation.
The events were extracted in the range between 200\,MeV and 300\,GeV within a 
$15^{\circ}$ acceptance cone of the Region of Interest (ROI) centered 
on the location of 3C~279 (RA = $195.047^{\circ}$, 
DEC=-$5.789^{\circ}$, J2000). Below 200\,MeV, the effective collection 
area of LAT for the diffuse class events drops very 
quickly and thus larger systematic errors are expected.  
The $\gamma$-ray flux and spectrum 
were calculated using the instrument response function (IRF) of ``P6\_V11\_DIFFUSE" 
by an unbinned maximum likelihood fit of model parameters. 
We examined the significance of the $\gamma$-ray signal from the sources
by means of the test statistic $(TS)$ based on the likelihood ratio 
test\footnote{$TS=25$ with 2 degrees of freedom corresponds to an estimated $\sim4.6\, \sigma$ pre-trials statistical significance assuming that the null-hypothesis $TS$ distribution follows a $\chi^2$ distribution \citep[see][]{ML}.}.
The background models included a component 
for the Galactic diffuse emission along the plane of the Milky Way, 
which was modeled by the map cube file 
``{\it gll\_iem\_v02\_P6\_V11\_DIFFUSE.fits}". An isotropic 
component ({\it isotropic\_iem\_v02\_P6\_V11\_DIFFUSE.txt}) 
was also included to represent the extragalactic diffuse emission 
and residual instrumental background.  Besides those components, 
the model in our analysis also included the emission from 
all nearby point sources inside the ROI from 
the first \Fermi-LAT\ catalog~\citep[1FGL:][]{1FGL}.
The spectra of those sources were 
modeled by power law functions 
except for a pulsar 1FGL~J1231.1--1410 (=PSR~J1231--1411),
for which we included an additional
exponential cut-off in its spectral modeling.
During the spectral fitting, 
the normalization factors of the Galactic diffuse and
isotropic components and the nearby sources were left as free 
parameters, and the photon indices of the nearby sources were fixed 
to the values from the 1FGL catalog except for 3C~273, 
whose photon index was allowed to vary freely.
In the light curve analysis, we considered only two 
bright sources in the background model as nearby point sources, namely 3C~273 and 1FGL~J1231.1--1410, 
because other nearby sources had a negligible contribution to 
$\gamma$-ray signal, especially in such relatively short time scales
(shorter than a week) for the light curves considered here.  
The fluxes used for the light curve 
were calculated by a simple power law model fit 
using data in the given energy ranges. 

\subsection{Temporal behavior}
\begin{figure*}[htpb]
  \centering
\includegraphics[width=12.9cm]{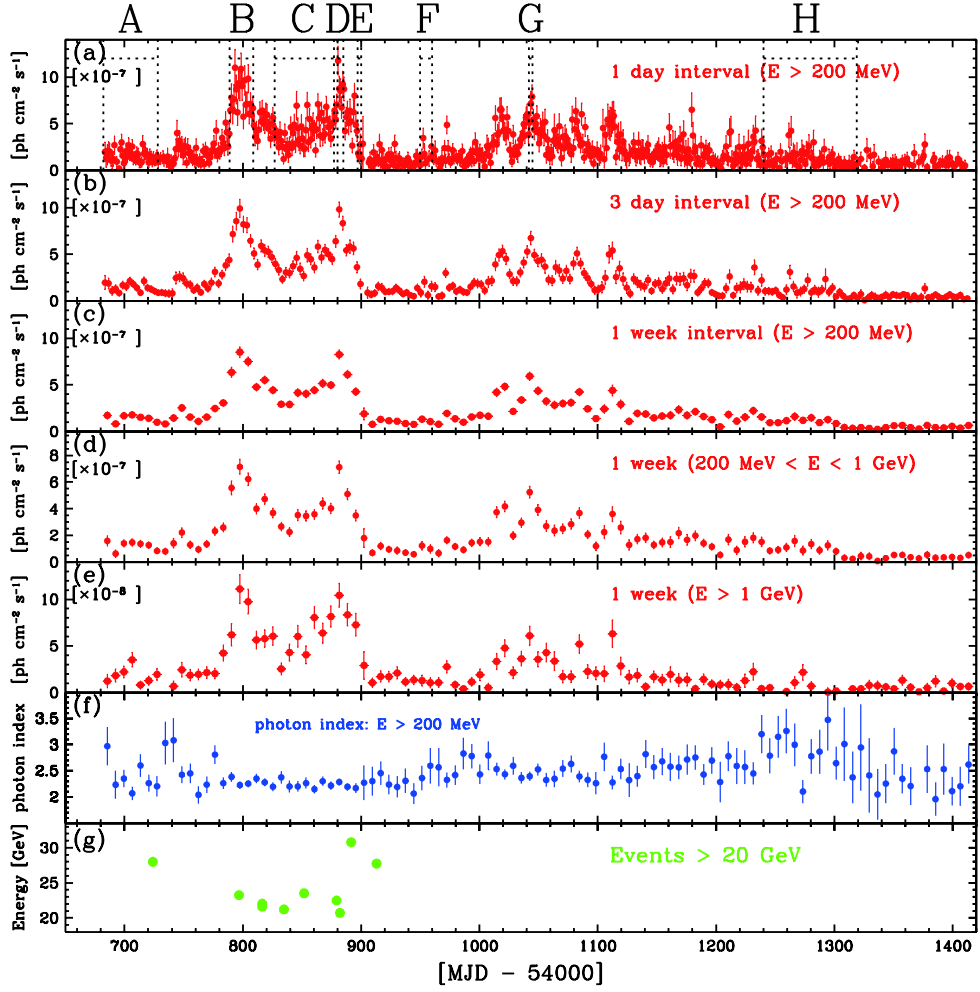}
\caption{Gamma-ray light curves of 3C 279 during the first two years of the \Fermi-LAT observations from 2008 August to 2010 August,
plotted in;  (a) 1-day intervals at energies above 200 MeV, 
(b) 3-day intervals at energies above 200 MeV, (c) 1-week intervals at  
energies above 200 MeV, (d) 1-week intervals at energies 
between 200 MeV and 1 GeV, (e) 1-week intervals at energies 
above 1 GeV.  The panel-(f) shows the history of the photon index 
at energies above 200 MeV in 1 week intervals, while the panel-(g) 
shows arrival time distribution of $>20$~GeV events associated with 3C~279. 
The vertical axis of the panel-(g) represents the estimated energy of events. 
The highest energy photon corresponds to 30.8 GeV at MJD~54891.
The dotted lines and capital letters represent time intervals where $\gamma$-ray spectra are extracted (see also Table~\ref{GammaSPfit}).}
  \label{GammaLC_time}
 \end{figure*}

The $\gamma$-ray light curve measured by \Fermi-LAT can be seen in 
Figure~\ref{GammaLC_time}.
The Figure shows the flux history above 200\,MeV 
averaged over (a) 1-day intervals, (b) 3-day intervals, 
and (c) 1-week intervals.  
It also includes 1-week light curves of (d) the flux 
between 200 MeV and 1 GeV, (e) the flux above 1 GeV, 
and (f) the photon index in the range above 200 MeV.

The $\gamma$-ray flux clearly shows variability. The source showed
high-flux states between MJD~54700 and 54900, in which two prominent 
flares can be seen: one of the flares at $\sim$ MJD~54800 and the the other at 
$\sim$ MJD~54880. During the second flare, a change in the optical 
polarization associated with a $\gamma$-ray flare was discovered 
(Paper I).  We detected some flux variability between 
MJD~55000 and 55120, but after that, the source remained in 
a relatively low-activity state until the end of the period considered 
in this paper. During this 2-year period, the highest 
integral flux above 200 MeV occured on MJD 54880 in the 
1-day interval light curve with flux of 
\ftwo$=(11.8\pm1.5)\times10^{-7}$ \phcms\ and $TS = 306$.
By extrapolating the spectrum down to 100 MeV,
an integral flux above 100~MeV on that day yields 
\fone$=(31.0\pm6.0)\times10^{-7}$ \phcms, 
which is still a factor 
of 3--4 times lower than the flux 
of the brightest flare ($\sim 1\times10^{-5}$ \phcms) 
detected during the EGRET observations of the source~\citep{Weh98, Har01vari}.

We quantified the flux variability using 1-week interval data
for energies above 200 MeV [full band], 
between 200 MeV and 1 GeV [soft band], and above 1 GeV [hard band].  
This is based on the ``excess variance'' method ~\citep{Nan97, Ede02} after subtracting 
the contribution expected from measurement errors ($\sigma_{{\rm err}, i}$).
Using the mean square error $<\sigma_{{\rm err}, i}>$, the excess variance $F_{\rm var}$  can be 
described as~\citep{Vau03}
 \begin{equation}
 F_{\rm var} =\sqrt \frac{S^2 - <\sigma_{{\rm err},i}>^2}{<F>^2} \label{eq:vari}
 \end{equation}
where $S$ is the variance of the flux, and $<F>$ is the mean value of 
the flux. The definition of associated error can be found in~\citep{Vau03}.
In the calculation, we excluded bins of 8, 82, 87 and 90 because the 
fit in the hard band failed due to poor statistics of the data samples.
Resulting $F_{\rm var}$ values are $0.695\pm0.015$,  $0.648\pm0.017$ 
and $0.839\pm0.030$ for the full, soft and hard bands, respectively.
The resulting values indicate that the flux of the hard band 
showed significantly stronger variability than that of the soft band.
For comparison, $F_{\rm var}=0.79\pm 0.02$ for $E>300\;{\rm MeV}$ has been reported during the first 11 months of the {\it Fermi} scientific mission~\citep{Variability}, when the source has clearly been more active.

\begin{figure}[htpb]
  \centering
\includegraphics[width=8.5cm]{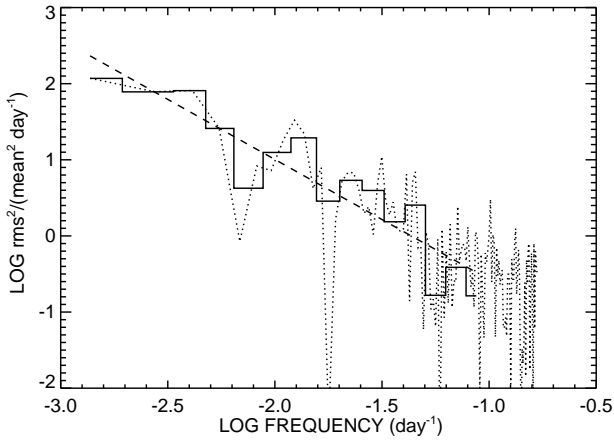}
\caption{Power Density Spectrum of 3C~279 for the 3-day binned $\gamma$-ray light curve. The white noise level has been subtracted.
The solid line histogram describes the PDS averaged in logarithmic frequency bins while the dotted curve describes the raw PDS before binning.
The dashed line represents a linear fit to the binned PDS.}.
  \label{PDS}
 \end{figure}

A Power Density Spectrum (PDS) for the 3-day binned light curve was calculated using a Fourier transform
and is shown in Figure~\ref{PDS}.
The power density was normalized to fractional variance per frequency unit
(\mbox{\it\,rms$^{2}$~I$^{-2}$~Day$^{-1}$}) and the PDS points were
averaged in logarithmic frequency bins.
The white noise level was estimated from the rms of the flux errors and was subtracted from the PDS.
A slope of $1.6\pm0.2$ was obtained from a linear fit to the binned PDS for frequencies up to 
0.1 \mbox{\it Day$^{-1}$}.
The main uncertainty in the estimated PDS slope is due to
the stochastic nature of the variability which leads to variations
in the determined slope between different time limited
observations. An additional effect which can cause a systematic
bias in the observed PDS slope is the red noise leakage~\cite[e.g.,][]{Cha08}.
In the present analysis this effect is not taken into account.

\begin{figure*}[tbp]
\center
\includegraphics[width=16.2cm]{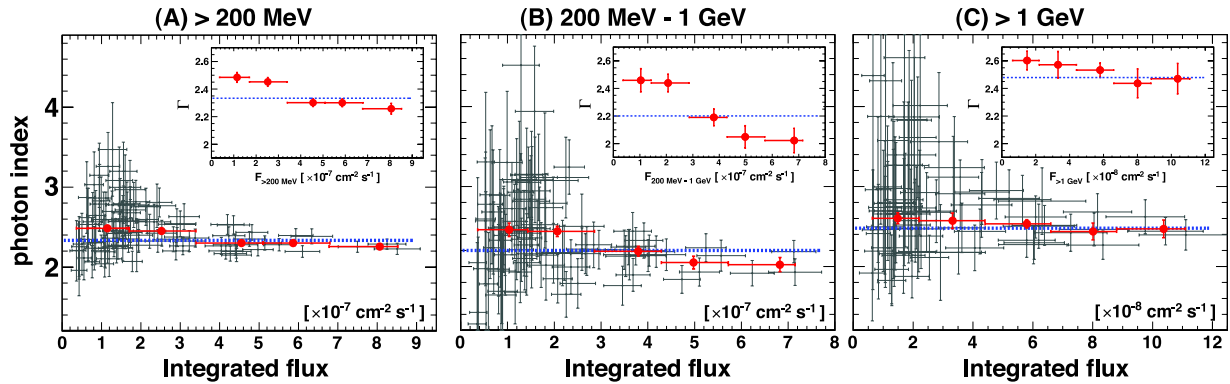}
  \caption{Plots of the integrated $\gamma$-ray flux vs.~photon 
index of 3C\,279 measured in week-long bins for energies above 200 MeV (A), between 
200\,MeV and 1\,GeV (B) and above 1\,GeV (C).
Only points with $TS > 10$ are plotted (gray points).
The blue dotted horizontal lines indicate average photon indices of those data for each energy band.
Red points show the photon indices resulting 
from an analysis where the data were sorted in five bins using week-long fluxes for each energy band.
For the red points, the horizontal bars indicate the ranges of the week-long flux bins
while the vertical bars indicate 1\,$\sigma$ statistical errors.
The insets show enlarged views of the red points as well as the average photon index of each energy band.
}
  \label{fig:FluxInd}
\end{figure*}

Figure~\ref{fig:FluxInd} shows plots of flux vs. photon index ($\Gamma$)
based on the weekly light curve results above 200\,MeV [full band],
between 200\,MeV and 1\,GeV [soft band] and above 1\,GeV [hard band].
The data which have $TS > 10$ were selected for the plots and are shown in gray points.
An average photon index was calculated by fitting a constant value in each plot,
corresponding to $\Gamma_{> 200\,{\rm MeV}} = 2.334\pm0.015$,
$\Gamma_{200\,{\rm MeV} - 1\,{\rm GeV}} = 2.20\pm0.03$ and 
$\Gamma_{> 1\,{\rm GeV}} =2.48\pm0.04$ for the full, soft and hard bands, respectively.
The average photon index in the soft band shows 
a significantly harder spectrum than that in the hard band.

We also derived photon indices resulting from an analysis
where the data were sorted in five bins using week-long fluxes for each energy band,
and plotted the results as red points.
Those photon indices of each flux bin are also shown in the insets in Figure~\ref{fig:FluxInd}.
For the full band,
although the change of the photon index is rather small ($\Delta\Gamma \sim 0.2$)
compared to the flux variation (spanning about an order of magnitude),
a weak ``harder when brighter'' effect can be seen.
Such an effect was also measured in other LAT blazars~\citep{LATSP}. 
The soft band also shows the weak ``harder when brighter'' effect 
with a slightly larger change of the photon index ($\Delta\Gamma \sim 0.4$).
On the other hand, the photon index of the hard band changes only sightly
($\Delta\Gamma \sim 0.1$) and is statistically consistent with a constant value.\\

\subsection{Highest energy photons}

During the 2-year observations,
the highest energy photon associated with 3C~279 was detected 
at MJD~54891.60745 with an estimated energy of 30.8~GeV.
The event was converted in the front-thin layers (so-called ``front event'') of the LAT detector
and still remains even when we apply the cleanest event selection 
(so-called ``data clean event''),
which was developed for studying extragalactic $\gamma$-ray background~\citep{EGB}.
The reconstructed arrival direction of the event is 
5\arcmin.7 (=0\degree.095) away from 3C~279, 
and is within the 68\% containment radius of the 
LAT PSF (0\degree.114\ in the IRF of ``P6\_V11\_DIFFUSE'') for front events at 30.8~GeV.
Based on our model fit of the epoch which contains that highest-energy photon, 
we find the probability that the photon was associated with 
3C~279 (as opposed to all other sources in the model including the
diffuse emission and nearby point sources) is 88.6\,\%.

In total, we found 10 events with estimated energies 
higher than 20\,GeV within an 0\degree.25 radius centered at 3C~279.
All events lay within a 95\% containment radius of the LAT PSF from 3C~279
and remain even after the ``data clean selection'' applied.
The number of expected background events above 20\,GeV within 
the 0\degree.25 radius at the location of 3C~279 for the 
2-year observations is only 0.16 events.  
The bottom panel of Figure~\ref{GammaLC_time} plots the  arrival time 
distribution of those 10 events.  All events except for two 
were detected between MJD~54780 and 54900 during the high activity states.
No photon above 20\,GeV associated with 3C~279 has been detected 
after MJD~54914 during the 2-year observations.

   
 \begin{figure}[htbp]
  \centering
\includegraphics[width=8cm]{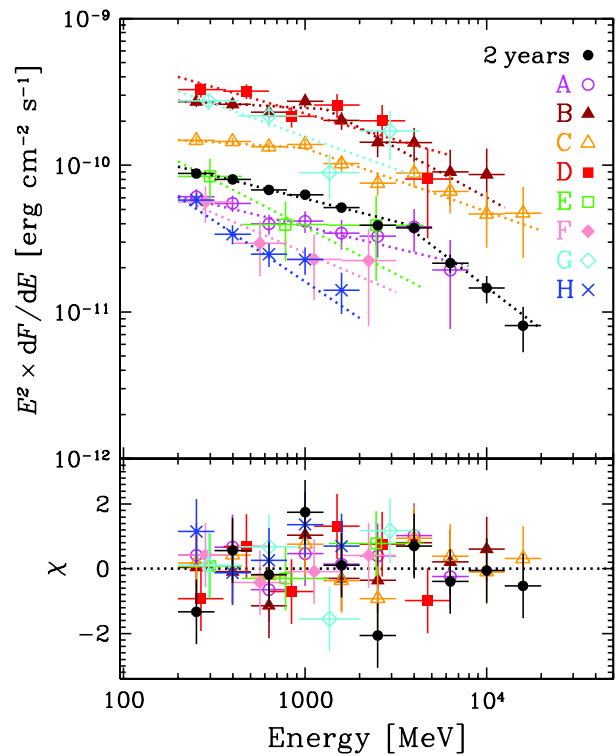}
  \caption{Gamma-ray spectral energy distributions of 3C~279 of each period as defined in the text or Table~\ref{GammaSPfit}. 2-year averaged [black:\,filled circles], Period A (magenta:\,open circles), Period B (brown:f\,illed triangles), Period C (orange:\,open triangles), Period D (red:\,filled squares), Period E (green:\,open squares), Period F (pink:\,filled diamonds), Period G (cyan:\,open diamonds) and Period H (blue:\,crosses). The vertical bars indicate $1\,\sigma$ statistical errors while the horizontal bars indicate energy ranges of each bin.
 The best-fit spectral models are plotted as dotted lines for each period and their parameters are summarized in Table~\ref{GammaSPfit}.
We use the broken power law model for the spectra of 2-year, Period B and Period C because significant improvements in the spectral fits can be seen compared 
to the simple power law model (see Table~\ref{GammaSPfit}) while the simple power law model is used for other periods. 
The lower panel shows the residuals, plotted as $\chi\,( \equiv ({\rm data} - {\rm model})/{\rm data\ error})$  from the best-fit models.
 A `dip' feature at $\sim$ 1--2~GeV in the spectrum of Period G (the 3rd point in cyan) is a $\le 2\,\sigma$ effect from the best-fit model, thus, not statistically significant.
  }
  \label{GammaSED}
 \end{figure}

\subsection{Gamma-ray spectra}
\label{sec:gamma_sp}

We extracted the $\gamma$-ray spectra using data for the 
entire 2-year period and following 8 sub-periods (see also Table~\ref{GammaSPfit}):  
[A] the initial quiescent state in the $\gamma$-ray band 
(MJD~54682 -- 54728), [B] the first $\gamma$-ray flaring state (MJD~54789 -- 54809), 
[C] an intermediate state (MJD~54827 -- 54877), [D] the first 
5 days of the second $\gamma$-ray flaring event (MJD~54880 --54885), 
[E] the last 3 days of the second $\gamma$-ray flaring event (MJD~54897 -- 54900), 
[F] during the isolated (first) X-ray flaring event (MJD~54950 -- 54960; see Section~\ref{section:results}), 
[G] during the second X-ray flaring event (MJD~55042 -- 55045; see Section~\ref{section:results}), 
and [H] a quiescent state (MJD~55240 -- 55319).
Those sub-periods were also selected taking into account observations in other energy bands.
Spectral energy distributions in the $\gamma$-ray band for each sub-period 
are presented in Figure~\ref{GammaSED}.
Each $\gamma$-ray spectrum was modeled using a simple 
power law (PL; ${\rm d}N/{\rm d}E \propto E^{-\Gamma}$), 
a broken power law (BPL; ${\rm d}N/{\rm d}E \propto E^{-\Gamma_1}$ 
for $E<E_{\rm brk}$ 
and ${\rm d}N/{\rm d}E \propto E^{-\Gamma_2}$ otherwise), 
and a log parabola (LogP; ${\rm d}N/{\rm d}E \propto (E/E_0)^{-\alpha - \beta \log(E/E_0)}$) model.
In the case of LogP model, the parameter 
$\beta$ represents the curvature around the peak. 
We note that the choice of the reference energy $E_0$ in the LogP 
model does not affect the determination of the other two model parameters, 
and hence we fixed it at 300 MeV.

\begin{deluxetable*}{c|ccccccc}
\tablecaption{Results of spectral fitting in the $\gamma$-ray band measured by \Fermi-LAT.\label{GammaSPfit}}
\tabletypesize{\scriptsize}
\centering                         
\startdata       
\hline 
Period & \multicolumn{6}{c}{Gamma-ray spectrum (\Fermi--LAT) } & Flux ($>100$ MeV) \\
(MJD) & fitting model\tablenotemark{a} & $\Gamma/\alpha/\Gamma_1$ & $\beta/\Gamma_2$ &   $E_{\rm brk}$ (GeV) &  $TS$ & $-2\Delta{L}$\tablenotemark{b} & $ (10^{-7}$ ph cm$^{-2}$ s$^{-1}$)  \\
\hline \hline
2 years &  PL & $2.38\pm0.02$ & \nodata & \nodata & 20272  & \nodata & $6.10\pm0.13$ \\
2008 Aug 4 -- 2010 Aug 6 &   LogP  & $2.18\pm0.03$ & $0.08\pm0.01$ & \nodata & 20267  & 46.5  & $5.18\pm0.16$ \\
(54682 -- 55414)   &  BPL & $2.31\pm0.02$ & $2.95\pm0.12$ & $3.5\pm0.3$ & 20286  & 43.0  & $5.76\pm0.15$ \\
\hline 
Period A  &  PL & $2.30\pm0.07$ & \nodata & \nodata & 797  & \nodata & $3.7\pm0.4$ \\
 2008 Aug 4 -- 2008 Sep 19 &   LogP  & $2.19\pm0.15$ & $0.04\pm0.05$ & \nodata & 797  & 0.7  & $3.3\pm0.5$ \\
(54682 -- 54728) &  BPL  & $2.21\pm0.09$ & $2.82\pm0.40$ & $3.4\pm0.8$ & 798  & 2.5  & $3.4\pm0.4$ \\
\hline 
Period B  &  PL & $2.28\pm0.04$ & \nodata & \nodata & 3209  & \nodata & $19.0\pm1.1$ \\
2008 Nov 19 -- 2008 Dec 9 &  LogP  & $1.95\pm0.10$ & $0.13\pm0.04$ & \nodata & 3214  & 13.6  & $15.0\pm1.3$ \\
(54789 -- 54809) &  BPL  & $2.00\pm0.10$ & $2.61\pm0.11$ & $1.0\pm0.2$ & 3215  & 13.7  & $15.8\pm1.3$ \\
\hline 
Period C  &  PL & $2.25\pm0.04$ & \nodata & \nodata & 4107  & \nodata & $10.0\pm0.5$ \\
2008 Dec 27 -- 2009 Feb 15 &  LogP  & $2.05\pm0.08$ & $0.08\pm0.03$ & \nodata & 4110  & 8.2  & $8.6\pm0.6$ \\
 (54827 -- 54877)   &  BPL  & $2.07\pm0.08$ & $2.43\pm0.08$ & $1.0\pm0.2$ & 4109  & 8.2  & $8.9\pm0.6$ \\
\hline 
Period D  &  PL & $2.36\pm0.08$ & \nodata & \nodata & 1236  & \nodata & $23.6\pm2.3$ \\
2009 Feb 18 -- 2009 Feb 23  &  LogP  & $2.16\pm0.16$ & $0.09\pm0.06$ & \nodata & 1234  & 2.0  & $20.2\pm2.8$ \\
  (54880 -- 54885) &  BPL  & $2.25\pm0.12$ & $2.91\pm0.61$ & $2.3\pm2.1$ & 1235  & 2.9  & $21.8\pm2.6$ \\
\hline 
Period E  &  PL & $2.64\pm0.32$ & \nodata & \nodata & 61  & \nodata & $6.3\pm2.5$ \\
 2009 Mar 7 -- 2009 Mar 10  & LogP  & $2.64\pm0.32$ & $0.00\pm0.00$ & \nodata & 61  & 0.0  & $6.3\pm2.5$ \\
(54897 -- 54900) \\ 
\hline 
Period F  &  PL & $2.54\pm0.24$ & \nodata & \nodata & 85  & \nodata & $3.5\pm1.2$ \\
2009 Apr 29 -- 2009 May 9&  LogP  & $2.54\pm0.24$ & $0.00\pm0.00$ & \nodata & 85  & 0.0  & $3.5\pm1.2$ \\
(54950 -- 54960)  \\
\hline 
Period G  &  PL & $2.44\pm0.13$ & \nodata & \nodata & 460  & \nodata & $18.8\pm2.9$ \\
 2009 Jul 30 -- 2009 Aug 2 &  LogP  & $2.37\pm0.25$ & $0.03\pm0.10$ & \nodata & 460  & 0.1  & $17.7\pm4.0$ \\
 (55042 -- 55045) \\
\hline 
Period H  &  PL & $2.83\pm0.11$ & \nodata & \nodata & 398  & \nodata & $3.7\pm0.5$ \\
 2010 Feb 13 -- 2010 May 3 &  LogP  & $2.56\pm0.23$ & $0.16\pm0.13$ & \nodata & 399  & 1.9  & $2.9\pm0.6$ \\
 (55240 -- 55319) &  BPL  & $2.72\pm0.43$ & $3.47\pm0.57$ & $1.6\pm0.5$ & 399  & 1.6  & $3.4\pm1.7$ 
\enddata
\tablenotetext{a}{PL: power law model, LogP: log parabola model, BPL: broken power law model. See definitions in the text.}
\tablenotetext{b}{ $\Delta L$ represents the difference of the logarithm of the likelihood of the fit with respect to a single power law fit.}
\end{deluxetable*}

The best-fit parameters calculated by the fitting procedure are 
summarized in Table~\ref{GammaSPfit}. The integral fluxes 
above 100\,MeV\footnote{Although we use photon data from 200 MeV, the integral fluxes are extrapolated down to 100 MeV, which
is convenient to compare with other $\gamma$-ray results.}
derived using each spectral model are also included.
The averaged $\gamma$-ray spectral shape for the 2-year observation
significantly deviates from a single power law.
A LogP model is favored to describe the $\gamma$-ray spectral shape over the simple PL model 
with the difference of the logarithm of the likelihood fits\footnote{$-2\Delta L = -2\log(L0/L1)$, 
where $L0$ and $L1$ are the maximum likelihood estimated for 
the null and alternative hypothesis, respectively.}
 $-2\Delta L =46.5$
(corresponding to a significance level of $\sim7\,\sigma$)\footnote{Because the LogP model has one more
free parameter than the PL model has,
 the $-2\Delta L$ distribution follows a $\chi^2$ distribution with 1 degree of freedom.},
and a BPL fit yields  $-2\Delta L =43.0$.
Even in some individual periods as defined 
above, the spectra deviate from a single power law: 
for example, the spectrum in the Period B yields
$-2\Delta L=13.6$.
This is consistent with our finding in Section 2.2 that the 
spectrum above 1 GeV is significantly softer than the spectrum below 1 GeV.
We thus conclude that the $\gamma$-ray spectrum significantly deviates from a simple power law.
The spectral break in 3C~279 is not as pronounced as that seen 
in the spectra e.g., of 3C~454.3~\citep{3C454a}.
One the other hand, the BPL model returns break energies within a few GeV range regardless of the flux levels
as observed in other bright FSRQs, such as 3C~454.3 and 4C+21.35~\citep{tan11}.
Such a spectral feature could be due to $\gamma-\gamma$ absorption to pair production
by He II Lyman recombination continuum UV photons from the emission line region~\citep[see, e.g.,][]{Pou10},
or a break in the electron distribution~\citep{3C454_2009}. 
We consider the $\gamma$-ray emission region to be located significantly beyond the 
broad emission region (see the Discussion in Section~\ref{section:modeling}), 
and this implies that the break in the electron energy distribution is the
more likely explanation.

\section{Multi-wavelength observations and data reduction}
\label{section:mwl}
\subsection{X-ray and Hard X-ray: \Suzaku}

The \Suzaku\ X-ray satellite~\citep{Mit07} observed 3C~279 as a part of multi-band studies of the 
object. 
The observations took place in two segments, 
with an interruption lasting roughly 1.5 days: 
(1) between 2009 January 19, 23:19:00 and 2009 January 22, 22:32:00 UTC (sequence number 703049010), 
and (2) between 2009 January 23, 20:45:00 and 2009 January 25, 03:00:00 UTC (sequence number 703049020).
``Period C" (see Table~\ref{GammaSPfit}) includes both \Suzaku\ observations.
The goals of the \Suzaku\ observations
were to monitor the soft-medium X-ray flux (0.3--12\,keV) of the source 
with the X-ray Imaging Spectrometer~\citep[XIS;][]{Koy07}
and to take advantage of the data from the 
Hard X-ray Detector~\citep[HXD;][]{Tak07}.
The HXD consists of PIN silicon diodes for the lower energy 
band (10--70~keV) and GSO scintillators 
for the higher energy band (40--600~keV),
to extend the spectral bandpass beyond the energies accessible 
with imaging instruments ($ >10$\,keV).  The HXD nominal 
position was used for the observations to maximize its effective area. In the following 
analysis, the HXD/GSO data were not used because there was no significant 
detection of the source.  

Although the observation conditions were nominal, the XIS1 data suffered 
from somewhat high and variable background, resulting in the 
total apparent counting rate ranging from 1 to 3 counts s$^{-1}$ 
in source-free regions for the entire chip.  Still, the background-subtracted 
spectrum determined from the XIS1 data below 8 keV was entirely consistent with 
that from XIS0 and XIS3 and thus we included the background-subtracted 
XIS1 data in the spectral fitting.  
The total duration of good data accumulated by the XIS instruments was 191 ks.  
We used the standard {\tt ftools} data reduction package, provided
by the \Suzaku\ Science Operations Center, with the calibration files 
included in CALDB ver.~4.3.1.  For the analysis of spectra and light 
curves, we extracted the counts from a region corresponding to a circle with 260\arcsec 
radius, centered on the X-ray centroid;  we used a region of a comparable 
size from the same chip to extract the background counts.  The net count rates 
were 0.47, 0.63, and 0.56 count s$^{-1}$ for XIS0, XIS1, and XIS3, respectively,
with the typical count rate uncertainty in the entire observations of $\sim 3\,\%$.   
The data indicate no significant variability during the \Suzaku\ 
observations.

The source was also detected in the HXD/PIN data, although the signal was relatively
weak.  We used the standard cleaned events, processed 
using the standard criteria applicable to the rev.~2.13 
of the \Suzaku\ HXD data processing software. This yielded 95.4 ks of good data, 
with a total count rate of 0.3 count s$^{-1}$.  For the background subtraction, 
we used the standard background files provided by the \Suzaku\ 
team through HEASARC.  We applied the standard tool {\tt hxdpinxbpi} which 
accounts for the particle background as well as for the contribution of the 
Cosmic X-ray Background as appropriate for the effective area and the solid 
angle of the HXD.  The net counting rate was 0.02 count s$^{-1}$, with the formal 
statistical uncertainty of $\sim 10$\%.  We note that this formal uncertainty is 
probably lower than the standard systematic error due to the background 
subtraction of 3\,\% of the average background (corresponding to 0.01 count s$^{-1}$).  
Nonetheless, even if the additional uncertainty is included, the source was still 
detected by the HXD/PIN.  

For the spectral analysis, we used the {\tt XSPEC} spectral analysis software.
For the spectral fitting of the XIS data, we used the standard 
redistribution files and mirror effective areas generated with \Suzaku-specific tools 
{\tt xisrmfgen} and {\tt xissimarfgen}.  In the spectral fits, we used 
the counts corresponding to the energy range of 0.5--10.0\,keV for XIS0 and XIS3, 
and 0.5--8.0\,keV for the XIS1.  We used all three XIS detectors simultaneously, 
but allowed for a small (a few \%) variation of normalization.  For the HXD/PIN data, 
we considered the data in the range of 20--50\,keV and used the response file 
{\tt ae\_hxd\_pinhxnome5\_20080716.rsp}.  

\begin{figure}[htbp]
  \centering
\plotone{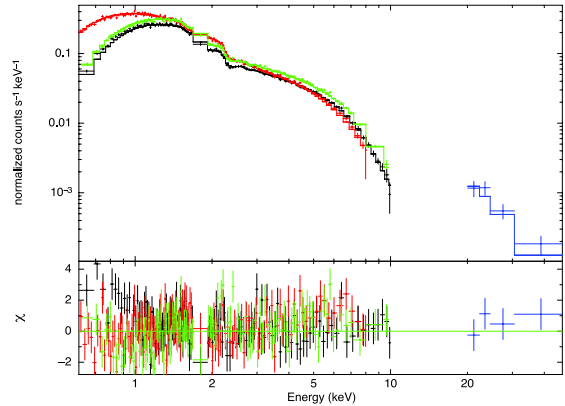}
  \caption{Count spectra of 3C~279 measured by \Suzaku~XIS0 (black), XIS1(red), XIS3 (green) and HXD/PIN (blue).
  The model plotted with the data is a broken power law obtained by a fitting these three XISs and HXD/PIN data.
  The lower panel shows the residuals for this broken power law model.}
  \label{SuzakuSP}
 \end{figure}

The source spectrum was modeled as an absorbed power law, with the 
cross-sections and elemental abundances as given in~\citet{Mor83};
other absorption models give similar results.  The best-fit 
absorbing column was $(3.1 \pm 0.5) \times 10^{20}$ cm$^{-2}$, 
and the photon index was $1.76 \pm 0.01$.  Inclusion of the HXD/PIN 
data in the fit did not change the fit parameters perceptibly.
The $\chi^{2}$ for the fit including the three XIS detectors and the HXD/PIN was 
acceptable, with 5061 for 5023 PHA bins.  The absorption inferred from 
the simple absorbed power law model is marginally greater than the value inferred 
from the radio measurements of the column density 
of the material in the Galaxy of $2.0 \times 10^{20}$ cm$^{-2}$ (with an estimated 
error of $\sim 10$\%) ~\citep{LAB}.
We deem the difference not significant, since at such small 
column densities, it can be accounted for by even small systematic uncertainty 
in the knowledge of the effective area of the XIS instruments at the lowest end 
of the XIS bandpass. 
Furthermore, a modest additional column density is expected in the host galaxy of 3C\,279.
The observed model 2--10 keV flux is $8.0 \times 10^{-12}$ 
erg cm$^{-2}$ s$^{-1}$, with the statistical error 
of $<2$\%, which is probably smaller than the systematic error resulting form the 
calibration uncertainty of the \Suzaku\ instruments.  
We plot the \Suzaku\ 3C~279 spectra in Figure~\ref{SuzakuSP}.  

   
\subsection{X-ray: \XMM}
\label{sec:xmm}
\XMM\ observed 3C~279 once starting on 2009 January 21, 17:28 UT.
The observation was largely devoid of flares (except for the 
period close to the end of the observation), and the total 
length of good data accumulated in the pointing was 16.8 ks.  
We used the standard Scientific Analysis System (SAS) data reduction package, provided
by the \XMM\ Science Operations Center.  Since 3C~279 is
a relatively bright source, we considered only the pn-CCD data.  
We note here that the spectra and light curves taken by MOS-CCDs were 
entirely consistent with the results inferred from the pn-CCD data. 

For the analysis of spectra and light curves, we extracted the 
counts from
within 40\arcsec radius of the source;  
we used a region of the same size, from the same pn-CCD chip, 
to extract the background counts.  The data indicate 
no significant variability during the \XMM\ observation. 
The spectral analysis was performed using the {\tt XSPEC}~v.12 spectral 
analysis software with the standard 
redistribution files and mirror effective areas included in the SAS 
package. We used the counts corresponding to the energy
range of 0.5--10.0 keV in our spectral fits.  

The source spectrum was first modeled as an absorbed power law;  
the best-fit absorbing column was $(2.2 \pm 0.6) \times 10^{20}$ cm$^{-2}$, 
and the photon index was $1.77 \pm 0.03$, with $\chi^{2}$ of 
588 for 517 d.o.f. The result is consistent with the spectral analysis results of 
the \Suzaku\ observations as described in the previous section, 
which were performed during the same period as the \XMM\ observation.
We also considered a broken power law model and found that 
the overall intrinsic source spectrum hardens with increasing energy.  
The absorbing column was $(3.4 \pm 0.7) \times 10^{20}$ cm$^{-2}$, and the low 
and high energy indices were respectively $1.83 \pm 0.05$ and $1.55 \pm 0.2$ 
with the break energy of $4.1 \pm 0.8$\,keV.  The resulting $\chi^{2}$ 
was 563, for 515 d.o.f.  
The broken power law model is statistically 
only marginally superior to the simple power law model,
especially given that the absorption inferred form 
the simple power law model is closer to the value inferred 
from~\citet{LAB}.
For either model, the 2--10\,keV flux is $7.7 \times 10^{-11}$ 
erg cm$^{-2}$ s$^{-1}$, with a statistical error 
of 5\%, which is probably smaller than the systematic error resulting form the 
calibration uncertainty of the \XMM\ pn-CCD.  


\subsection{X-ray: \RXTE-PCA}
\RXTE~carried out 321 observations between 2008 July 3rd (MJD~54650) 
and 2010 August 12th (MJD~55420).  
Those include 52 observations based on 
the Cycle 12 Guest observer (GO) program and 269 observations 
based on the Core program in Cycle 12--14. The fluxes resulting from  
the Cycle12 GO observations have been already reported in Paper I.
Most of the individual observations have exposure times in a range from 1.0 to 2.5\,ks.
We analyzed the data from the Proportional Counter Array (PCA) following standard 
procedures using the {\tt rex} script in HEASOFT~v.6.9.
Only signals from the top layer (X1L and X1R) of PCU2 were extracted for data analysis.
The data were screened with the 
following data selection: source elevation above the horizon $>$ 10$^\circ$, pointing 
offset smaller than 0.02$^\circ$, at least 30 minutes away from a SAA passage 
and electron contamination smaller than 0.1. 
The background was estimated with standard procedures,
and the detector response matrices were
extracted with the RXTE tools (command PCARSO~v.11.7.1). 
For the spectral analysis we re-binned the spectra into 11 channels. 
The spectra from the channels corresponding to nominal energies of 
2.6 to 10.5 keV are adequately fitted by a single power law model, absorbed 
by a fixed Galactic column density of $2.2\times10^{20}$ cm$^{-2}$
using the {\tt XSPEC}~v.12 software package.
The value of the column density is based on our \XMM\ results (in Section~\ref{sec:xmm}) 
and is also consistent with the value based on \citet{LAB}.

\subsection{X-ray: \Swift-XRT}
In the HEASARC data base\footnote{\url{http://heasarc.gsfc.nasa.gov/cgi-bin/W3Browse/swift.pl}},
there are 80 publicly available \Swift\ X-Ray Telescope (XRT) observations between 
2008 July 3 (MJD~54650) and 2010 August 12 (MJD~55420),
which include 32 pointings based on an approved GI proposal in Cycle-4 (Proposal number: 5080069).
The results of the flux history based on the data until 2009 May 31 
have already reported in Paper I.  
Effective exposure times of these observations
range between 1 and 3 ks, but some have longer exposure times, for example,  
 8.9 ks for ID:35019007 (MJD~54795), 22.5 ks for ID:35019009 (MJD~54797), 
 20.2 ks for ID:35019010 (MJD~54799) and 15.4 ks for ID:35019011 (MJD~54800).
The XRT was used in the photon counting mode, and no evidence of pile-up was found. 
The XRT data were reduced with the standard software {\tt xrtpipeline} 
v.0.12.6, applying the default filtering and screening criteria (HEADAS
package, v.6.10).  
The source events were 
extracted from a circular region, 20 pixels in radius, centered on the source position.
Exposure maps were used to account for
Point-Spread Function losses and the presence of dead pixels/columns.
The background was determined using data 
extracted from a circular region, 40 pixels in radius, centered on 
(RA, Dec: J2000) =(12h56m26s,-05\degree49\arcmin30\arcsec), where no X-ray sources are found.
Note that the background contamination is less 
than 1\,\% of source flux even in the faint X-ray states of the source.
The data were rebinned to have at least 25 counts per bin, and the spectral 
fitting was performed using the energy range between 0.3 keV and 10 keV using {\tt XSPEC}~v.12.
The Galactic column density is fixed at $2.2\times10^{20}$ cm$^{-2}$ during the fittings
as is the case in the \RXTE\ data analysis.

\begin{figure}[htbp]
  \centering
\includegraphics[width=5.3cm,angle=-90]{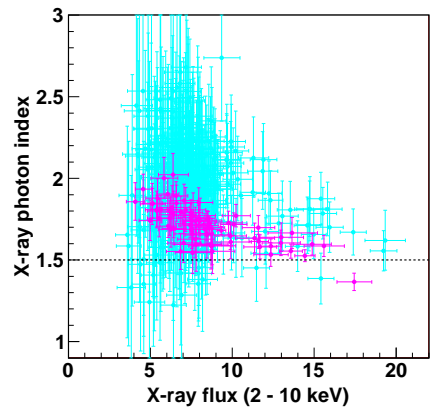}
  \caption{Scatter plot of flux vs.~photon index of 3C~279 in the 
   X-ray band with the data taken by \Swift-XRT (magenta) 
   and \RXTE-PCA (cyan). The horizontal dotted line represents 
   the photon index value of 1.5.}
  \label{X_FvsI}
\end{figure}

Figure~\ref{X_FvsI} shows a scatter plot between photon index and flux in the 
X-ray band as measured by \Swift-XRT and \RXTE-PCA.
Generally, a ``harder-when-brighter'' trend can be seen. 
Only the highest-flux point measured by \Swift-XRT shows the photon index significantly harder (smaller) than 1.5.

\subsection{Ultra-Violet: \Swift-UVOT}

\begin{figure}[htpb]
  \centering
 \includegraphics[height=9.7cm]{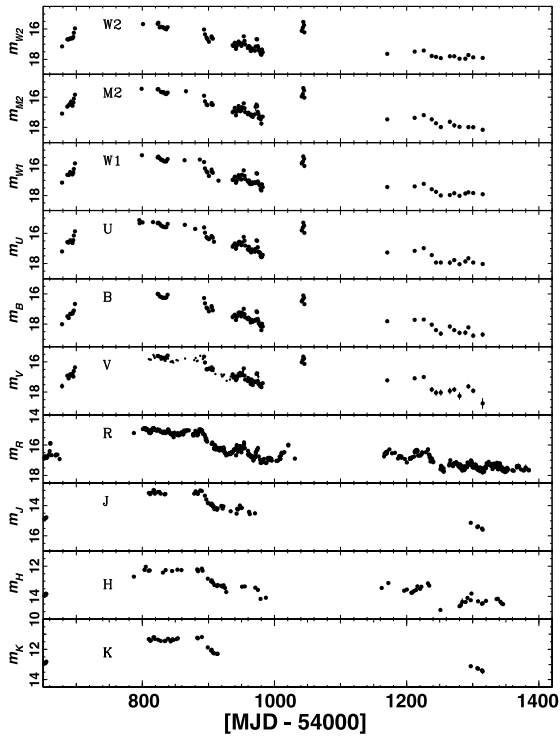}
\caption{Light curves of all observed UV-optical-NearIR bands of 3C~279 in measured magnitude scale from 2008 August to 2010 August, including $W2$ (\Swift-UVOT), $M2$ (\Swift-UVOT), $W1$ (\Swift-UVOT), $U$ (\Swift-UVOT),  $B$ (\Swift-UVOT), $V$ (Katana, \Swift-UVOT), $R$ (Abastumani, Calar Alto, ST-7, GRT, MDM, L'Ampolla, Perkins, SLT, KVA, LT, San Pedro, St.~Petersburg, Tijarafe),  $J$ (AZT-24, Kanata), $H$ (AZT-24, LT) and $K$ (AZT-24, Kanata) bands.}
  \label{optiLC}
 \end{figure}

The \Swift\ Ultra-Violet/Optical Telescope (UVOT; Roming et al. 2005) data used in
this analysis included all of the observations performed during 
the time interval MJD 54650--55420.
The UVOT telescope cycled through each of the six optical 
and ultraviolet filters $(V, B, U, W1, M2, W2)$.
The UVOT photometric system is described in \citet{key:poole2008}.
Photometry was computed from a 5\arcsec\ source region around 3C~279
using the publicly available UVOT {\tt FTOOLS} data
reduction suite.
The background region 
was taken from an annulus with inner and outer radii of 27\arcsec.5 and 35\arcsec, respectively. 
Galactic absorption in the direction of 3C 279 was adapted as given in~\citet{Lar08}, 
namely, $A_V = 0.093$, $A_B=0.123$, $A_U=0.147$, $A_{W1}=0.195$, $A_{M2}=0.285$ and $A_{W2}=0.271$.
The measured magnitudes in each band during the 2-year observations are
$m_V = 15.6 - 18.7$ (75 data points), $m_B = 16.0 - 18.8$ (80 data points), $m_U = 15.1 - 18.0$ (88 data points), $m_{W1}=15.3 - 18.0$ (84 data points), $m_{M2} = 15.4 - 18.2$ (76 data points) and $m_{W2}=15.5-18.0$ (81 data points).
 All observed data points are shown in Figure~\ref{optiLC}.
 
\subsection{Optical, Near-Infrared and Radio observations by GASP-WEBT}

The GLAST-AGILE Support Program (GASP; \citealt{vil08,vil09}) is a project initially
originating from the 
Whole Earth Blazar Telescope\footnote{\url{http://www.oato.inaf.it/blazars/webt}} (WEBT) in 2007.
It is aimed to provide long-term monitoring in the optical ($R$ band), near-IR, and mm--cm radio 
bands of 28 $\gamma$-ray-loud blazars during the lifetime of the AGILE and \Fermi\ $\gamma$-ray 
satellites.  

The observations of 3C~279 in the period considered in this paper were performed by the 
observatories listed in Table~\ref{telescope}.  The calibrated $R$-band magnitudes of 
the source were obtained through differential photometry with respect to the reference 
stars 1, 2, 3, and 5 by \citet{rai98}. Near-IR data in the $J$, $H$, and $K$ filters 
were acquired at Campo Imperatore and Roque de los Muchachos (Liverpool).
When converting magnitudes into flux densities, optical and near-IR data 
were corrected for Galactic reddening using $A_B=0.123$ mag \citep{sch98}.
We adapted the extinction laws by \citet{car89}, and the zero-mag fluxes 
by \citet{bes98}. 

\begin{figure}[tpb]
  \centering
 \includegraphics[height=9.7cm]{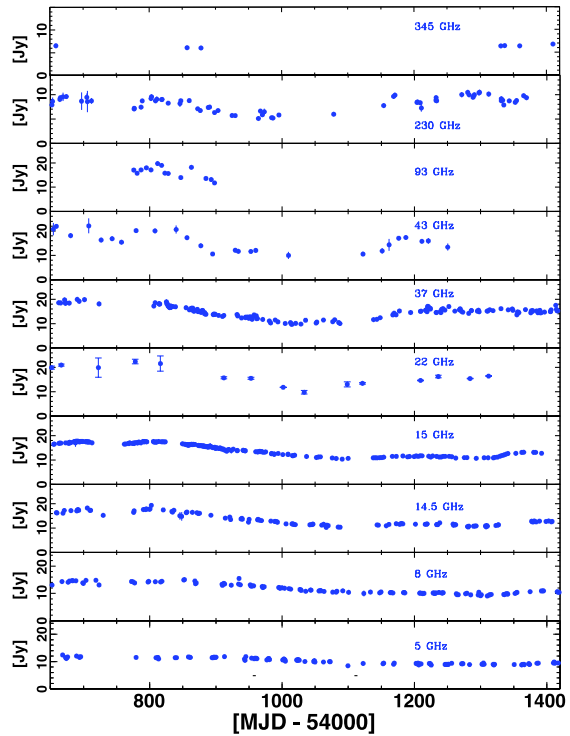}
\caption{Radio light curves of 3C~279 from 2008 August to 2010 August measured at all observed radio frequencies: 345\,GHz (SMA), 230\,GHz (CARMA, SMA), 93\,GHz (CARMA), 43\,GHz (Noto), 37\,GHz (Metsahovi), 22\,GHz (Medicina), 15\,GHz (OVRO), 14.5\,GHz (UMRAO), 8\,GHz (Medicina, UMRAO) and 5\,GHz (Medicina, UMRAO).}
  \label{radioLC}
 \end{figure}

For the observations between 2008 August and 2010 August, 
the measured $R$-band magnitude ranged
from 14.87 to 17.81 (673 data points).
The $R$-band data have the best time coverage among the IR-optical-UV bands in our data
thanks to the participation of a number of telescopes.
The emission shows strong variability and
the excess variance ($F_{\rm var}$; Eq.~\ref{eq:vari}) of the source $R$-band flux (i.e., in linear scale) is $0.853\pm0.001$.
The near-IR magnitudes in the $J$, $H$ and $K$ bands were measured in ranges of 
$m_J = 14.91 - 15.59$ (20 data points), $m_H=12.04-14.90$ (68 data points)
and $m_K=11.19-13.45$ (20 data points).
Those data points are shown in Figure~\ref{optiLC}.
The radio flux densities were measured in ranges of
$F_{\rm 5\,GHz}=8.5 - 12.4$ Jy (109 data points), 
$F_{\rm 8\,GHz}=9.1 - 15.5$ Jy (124 data points), 
$F_{\rm 14.5\,GHz}=10.3 - 19.4$ Jy (118 data points), 
$F_{\rm 22\,GHz}=10 - 22 $ Jy (16 data points), 
$F_{\rm 37\,GHz}=10 - 20 $ Jy (168 data points), 
$F_{\rm 43\,GHz}=10 - 22 $ Jy (20 data points),
$F_{\rm 230\,GHz}=5.1 - 10.5 $ Jy (62 data points)
and $F_{\rm 345\,GHz}=6.0 - 6.8 $ Jy (7 data points).
The light curves of the radio flux densities in those bands are plotted in Figure~\ref{radioLC}.

\begin{deluxetable*}{lcc} 
\tabletypesize{\scriptsize}
\tablecaption{List of observatories that contributed data to the campaign}
\vspace{3mm}
\centering
\startdata
\hline
Observatory & Detector/Telescope (diam.) &   Band \\ 
\hline \hline
\multicolumn{3}{c}{{\it Gamma ray}} \\

\Fermi  & LAT (survey mode) & $>200$ MeV \\
\hline
\multicolumn{3}{c}{{\it X ray}} \\
\Suzaku & XIS & 0.5 -10 keV\\
 & HXD/PIN & 15 - 50 keV \\
 \XMM & PN & 0.5 -10 keV \\
\RXTE & PCA & 3-10 keV  \\
\Swift & XRT & 0.6 - 7 keV  \\
\hline
\multicolumn{3}{c}{{\it Ultra-Violet, Optical, Infrared}} \\
\Swift & UVOT & $W2, M2, W1, U, B, V $ \\
\Spitzer &  IRS & 5 - 38 $\mu$m\\
 & IRAC & 3.6, 4.5, 5.8, 8.0 $\mu$m \\
  & MIPS & 24, 70, 160 $\mu$m \\
Abastumani, Georgia$^{\#}$ &  (70 cm)  & $R$  \\  
Calar Alto$^{\#, *}$ &  & $R$  \\
Campo Imperatore, Italy$^{\#}$ & AZT-24 (110 cm) &  $J, H, K$ \\
Crimean, Ukraine$^{\#}$ & ST-7 (70 cm)  & $R$  \\
Goddard, USA$^{\#}$ & GRT & $R$ \\
Hiroshima, Japan  & Kanata (150 cm)  &  $V, J, Ks$, polarization ($V$)  \\
Kitt Peak, Arizona, USA$^{\#}$ & MDM (130 cm) & $R$  \\
La Silla, Chile & GROND (220 cm)  & $g, r, i, z, J, H, K$ \\
L'Ampolla$^{\#}$ &  & $R$  \\
Lowell (Perkins)$^{\#}$ & Perkins  & $R$  \\
Lulin, Taiwan$^{\#}$ & SLT (40 cm)  & $R$  \\
Roque, Canary Islands$^{\#}$ &  KVA (35 cm)  & $R$, polarization (no filter)  \\
Roque, Canary Islands$^{\#}$ & LT (200 cm) & $R, H$  \\
San Pedro Martir$^{\#}$ &  (84 cm)  & $R $ \\
St. Petersburg, Russia$^{\#}$ & (40 cm) & $R $ \\
Tijarafe$^{\#}$ & (35 cm) & $R$ \\
\hline
\multicolumn{3}{c}{{\it Radio}} \\

CARMA, USA & (array)  & 92.5, 227.5 GHz \\
Mauna Kea, USA$^{\#}$ &  SMA($8\times6$~m)   & 230, 345 GHz  \\
Medicina, Italy$^{\#}$  & (32 m) & 5, 8, 22 GHz \\
Metsahovi, Finland$^{\#}$ & (14m) & 37 GHz \\
Noto, Italy$^{\#}$ & (32 m) & 43 GHz  \\
Owens Valley, USA & OVRO (40~m) & 15 GHz \\
UMRAO, USA$^{\#}$ & (26 m) & 5, 8, 14.5 GHz 
\enddata
\tablenotetext{\#}{GASP-WEBT}
\tablenotetext{*}{Calar Alto data was acquired as part of the MAPCAT project:http://www.iaa.es/~iagudo/research/MAPCAT}
\label{telescope}

\end{deluxetable*}


\subsection{Optical and Near-Infrared: the Kanata telescope}

We performed the $V$, $J$ and $Ks$-band photometry and polarimetry
of 3C 279 using TRISPEC installed to the 1.5\,m Kanata telescope located in the
Higashi-Hiroshima Observatory.

TRISPEC has a CCD and two InSb arrays, enabling photopolarimetric
observations in an optical and two near-IR bands simultaneously~\citep{wat05}.
We obtained 64, 42 and 17 photometric measurements in the {\it V, J} and {\it Ks} bands, respectively.
A unit of the polarimetric observing sequence consisted of successive
exposures at 4 position angles of a half-wave
plates:  $0^{\circ},45^{\circ},22^{\circ}.5,67^{\circ}.5.$
The data were reduced according to the standard procedures of CCD photometry.
We measured the magnitudes of objects with the aperture photometry technique.
We performed differential photometry with a comparison star taken in the
same frame of 3C 279. Its position is R.A.=12h56m16.90s,
Dec=-05\degree50\arcmin43.0\arcsec (J2000)
and its magnitudes are {\it V} = 13.660, {\it J} = 12.377 and {\it Ks} =11.974~\citep{rai98, cut03}. 
The photometric data have been corrected for the Galactic extinction
with $A_V = 0.093$, $A_J = 0.026$, and $A_{Ks} = 0.010$. 
The measured optical and near-IR magnitudes by Kanata in the $V$, $J$ and $Ks$ bands
during the 2-year observations spanned $m_V = 15.54 - 17.27$ (56 data points),
 $m_J = 13.00 - 14.58$ (37 data points) and  $m_{Ks} = 11.21 - 11.47$ (17 data points).
Those data points are also shown in Figure~\ref{optiLC}.

We confirmed that the instrumental polarization was smaller than 0.1\%
in the {\it V} band using the observations of unpolarized standard stars. 
Hence, we did not apply any corrections for it.
The zero point of the polarization
angle is corrected as standard system (measured from north to east) by
observing the polarized stars, HD19820 and HD25443~\citep{wol99}.
The polarization shows clear variability and 
the degree of polarization was measured in the range of 3 -- 36\,\% 
during our 2-year observational campaign.
As we reported in Paper I,
we found a rotation of the polarization angle by 208\degree\ 
together with a sharp drop of the degree of polarization from $\sim 30$\,\% down to a few \%.
The event was coincident with a $\gamma$-ray flare (Period~D-E).
In the second half of the 2-year observations,
the source was generally in a quiet state in the optical band, and the degree of polarization was also relatively low.\\

\subsection{Optical and Near-Infrared: GROND}
The Gamma-Ray burst Optical/Near-Infrared Detector (GROND;~\citealt{Gre08}) 
mounted at the MPI/ESO 2.2~m telescope at 
LaSilla observatory in Chile observed the field of 3C~279 in two nights 
of 2008 July (2008 July 30 and 2008 July 31) and four nights in 2009 January
(2009 January 19 to 2009 January 22). In each observation, a total of 4 images in 
each $g^\prime  r^\prime i^\prime z^\prime$ filter with integrations 
times of 35~s and 24 images of 10~s exposure in each $JHK_s$ were 
obtained simultaneously.

GROND optical and near-IR data were reduced in standard manner using 
pyraf/IRAF \citep{1993ASPC...52..173T} similar to the procedure outlined 
in \citet{2008ApJ...685..376K}. The stacked images of each observation 
were flux calibrated against GROND observations of SDSS fields 
\citep{2009ApJS..182..543A} taken immediately before or after the field 
of 3C~279 for the optical $g^\prime  r^\prime i^\prime z^\prime$, and 
magnitudes of 2MASS field stars~\citep{2006AJ....131.1163S} for the 
$JHK_s$ filters. All data were corrected for the expected Galactic 
foreground reddening of $E_{(B-V)}=0.029$ according to \citet{sch98}.
Results of GROND observations are summarized in Table~\ref{GRONDobs}.

 \begin{deluxetable*}{c|cc|cc|c}
\tablecaption{Results of GROND observations.\label{GRONDobs}}
\centering                         
\startdata       
\hline
& \multicolumn{2}{c|}{2008 July 31}  &  \multicolumn{2}{c|}{2009 January 19-22}  & \\ \hline
filter &  AB magnitude & flux [mJy] & AB magnitude & flux [mJy] & $A/A(V)\tablenotemark{a}$ \\ \hline
$g^\prime$ &	$17.62\pm0.05$ & 	 $0.324\pm0.015$ &	  $16.06\pm0.05$ &	  $1.37\pm 0.06 $ & 1.23 \\
$r^\prime$&	$17.24\pm0.05  $ & 	 $ 	 0.462\pm0.021	$ & 	 $  15.49\pm0.05$ & 	 $	   2.31 \pm 0.11 $ & 0.80 \\
$i^\prime$&	$16.83\pm0.05	$ & 	 $ 0.671\pm0.031$ & 	 $	  15.05\pm0.05	$ & 	 $   3.47 \pm0.16$ & 0.62 \\
$z^\prime$&	$16.67\pm0.05	$ & 	 $ 0.776\pm0.036$ & 	 $	  14.80\pm0.05	$ & 	 $   4.37\pm0.20$  & 0.45 \\
$J$&	 $16.05\pm0.06$ & 	 $	 1.387\pm0.077$ & 	 $	  14.16\pm0.06	$ & 	 $   7.86\pm0.44$ & 0.29 \\
$H$&	$15.60\pm0.07$ & 	 $	 2.098\pm0.133$ & 	 $	  13.62\pm0.07$ & 	 $	 12.90\pm0.81$ & 0.18 \\
$K$&	$15.18\pm0.09	$ & 	 $ 3.062\pm0.249$ & 	 $	  13.19\pm0.09	$ & 	 $ 19.18\pm1.56$ & 0.14 
\enddata
\tablenotetext{a}{dereddening factors for correction of Galactic extinction.}
\tablecomments{Results of both AB magnitude and flux are corrected for Galactic extinction. No significant daily variability was observed 
during the observations in 2009 January.}
\end{deluxetable*}

\subsection{Infrared: \Spitzer}
\label{sec:Spitzer}

We observed 3C~279 with \Spitzer\ Infrared Spectrograph (IRS), Multiband Imaging Photometer for Spitzer (MIPS) and Infrared Array Camera (IRAC) at several epochs in 2008 and 2009 under 
the \Spitzer\ program PID50231 (PI A.~Wehrle; see Table~\ref{Spitzer_log}).
The observations were conducted once with each instrument in 2008 July and August and
approximately daily during the instrument campaigns in the 2009 February--March
visibility window.

\begin{deluxetable*}{c|lc|c|c}
\tablecaption{\Spitzer\ observation log.\label{SpitzerLog}}
\centering                         
\startdata       
\hline 
Instrument	&	Start time (UTC)		& (MJD)	&	Duration (min)	&	ObsID	\\
\hline \hline
MIPS	&	2008.7.31 11:09:06.3	& 54678.4647	&	9.19	&	27434240	\\
	&	2009.2.15 07:08:08.6	& 54877.2973	&	14.61	&	27438592	\\
	&	2009.2.16 23:12:32.0	& 54878.9670	&	14.61	&	27438080	\\
	&	2009.2.17 20:42:36.5	&54879.8629	&	14.61	&	27438848	\\
	&	2009.2.18 19:39:30.8	& 54880.8191	&	14.60	&	27438336	\\
	&	2009.2.19 14:19:04.4	&54881.5966	&	14.60	&	27439360	\\
	&	2009.2.20 09:23:37.4	& 54882.3914	&	14.60	&	27439104	\\
\hline
IRS	&	2008.8.16 14:32:42.6	&54694.6060	&	17.92	&	27425024	\\
	&	2009.3.3 12:22:35.7	&54893.5157	&	18.15	&	27435776	\\
	&	2009.3.4 19:53:03.5	&54894.8285	&	18.15	&	27437312	\\
	&	2009.3.6 10:50:02.7	& 54896.4514	&	18.16	&	27436544	\\
	&	2009.3.7 00:46:03.4	& 54897.0320	&	18.16	&	27435520	\\
	&	2009.3.8 12:08:53.5	& 54898.5062	&	18.18	&	27437056	\\
	&	2009.3.9 18:39:51.2	& 54899.7777	&	18.18	&	27436288	\\
\hline
IRAC	&	2008.8.17 06:57:26.3	& 54695.2899	&	10.64	&	27429632	\\
	&	2009.3.10 19:26:05.5	& 54900.8097	&	10.69	&	27433216	\\
	&	2009.3.11 22:05:54.7	& 54901.9208	&	10.70	&	27432448	\\
	&	2009.3.13 05:13:56.9	&54903.2180	&	10.71	&	27433728	\\
	&	2009.3.14 04:12:41.8	& 54904.1755	&	10.72	&	27432960	\\
	&	2009.3.15 03:04:16.9	&54905.1280	&	10.73	&	27433984	\\
	&	2009.3.16 14:58:46.6	& 54906.6241	&	10.75	&	27433472	
\enddata
\label{Spitzer_log}
\end{deluxetable*}

\begin{figure}[htbp]
  \centering
 \includegraphics[height=7.5cm, angle=270]{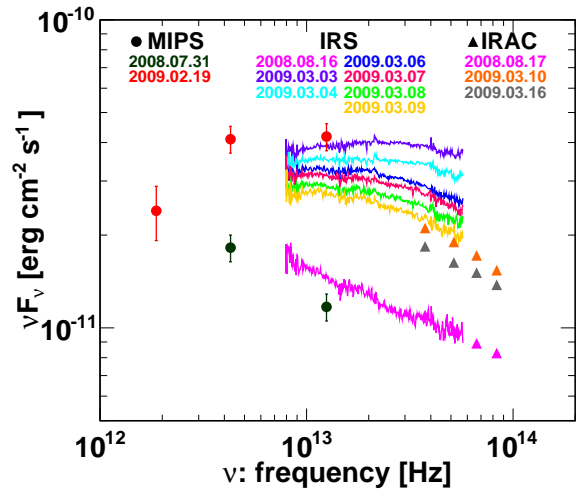}
  \caption{Spectral energy distribution of 3C~279 in the infrared band measured by \Spitzer-IRS. 
  IRS spectra from highest to lowest are on 2009 March 3, 2009 March 4, 
  2009 March 5, 2009 March 6, 2009 March 7, 2009 March 9, and 2008 August 16.
  Representative flux data points measured by \Spitzer-MIPS (circles) and \Spitzer-IRAC (triangles) are also included.}
  \label{IRSP}
 \end{figure}

For the IRS observations, high-accuracy blue peakup observations on a nearby star were used to
center the spectrograph slit on the target. 3C~279 was observed with the low-resolution 
SL2, SL1, LL2, and LL1  modules, for 3 cycles of 14 seconds at each of two nod positions.  
\Spitzer-IRS data reductions began with S18.7 \Spitzer\ Science Center pipeline-processed, 
background subtracted data.  The background was removed by subtracting the alternate 
nod for each pointing.  Additional processing steps were applied to clean bad data, 
remove fringes, and match and trim spectral orders. 
First, we cleaned bad, rogue pixels 
using the \Spitzer\ Science Center 
procedure IRSCLEAN V2.0. One-dimensional spectra were then extracted using 
the standard point-source aperture and flux calibration in SPICE ver.~2.3. We used 
a custom spectral defringing tool to remove fringes introduced by the pointing-dependent 
instrumental flatfield. This tool uses a predetermined flatfield fringing correction 
function, which is shifted to match and remove the observed fringes in the spectrum. 
Spectral orders were trimmed, and the SL2 and SL1 orders were scaled up by a factor of 
1.06 to empirically correct for pointing-dependent point-source slit losses. 
Finally, the nod-spectra were averaged and combined into a single spectrum 
covering 5.2--35\,$\mu$m rest wavelength.
Figure~\ref{IRSP} shows the reduced IRS spectra in the $\nu F_\nu$ representation.

We used the pipeline MIPS images (ver.~18) for aperture photometry using 
$13''$, $35''$ and $50''$ radius for 24, 70 and 160 $\mu$m bands, respectively, with 
aperture corrections from Tables 3.13, 3.14 and 3.16 of the MIPS Data Handbook 
(ver.~3.2). No 160 $\mu$m data were obtained in August 2008 because the array was 
not cooled during that campaign.  3C~279 has very low ecliptic latitude ($0.2^\circ$), 
hence, the observed transients can be attributed to passing asteroids 
which appeared in various MIPS images.
We used $6''$ radius apertures for IRAC photometry on the pipeline data (ver.~18) with
aperture corrections tabulated in Table 5.7 of the IRAC Data Handbook (ver.~3.0).

\begin{figure}[tbp]
  \centering
    \vspace{1.5mm}
 \includegraphics[width=7.2cm]{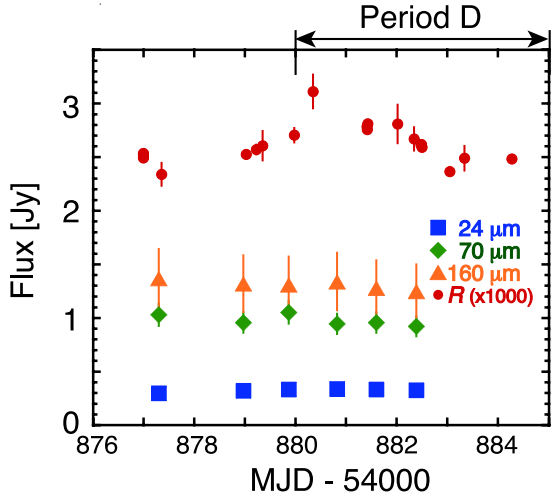} 
  \caption{Light curve of 3C~279 at $24\,\mu$m, $70\,\mu$m and $160\,\mu$m measured by Spitzer-MIPS. 
  The error bars correspond to 10\%, 10\% and 20\%, respectively, 
  for each band as mentioned in the text in Section~\ref{sec:Spitzer}.
  Optical $R$-band data taken by the ground-based telescopes are also 
  plotted in red color for comparison.}
  \label{IRLC}
 \end{figure}

The \Spitzer-MIPS photometric repeatability and absolute calibration 
uncertainties at 24\,$\mu$m are respectively, 0.4\,\% and 4\,\%; at 70\,$\mu$m, 
4.5\,\% and 5\,\%; and at 160\,$\mu$m, 5\,\% and 12\,\% ~\citep{eng07, gor07, sta07}.
We therefore adopt overall uncertainties of 10\,\%, 10\,\% and 20\,\% at 24, 70 and 160\,$\mu$m,
respectively.  No color correction has been applied to the data because 
the slope and smoothness of the spectrum over the bandpasses are not known.
Figure~\ref{IRLC} describes the MIPS flux history during the 6 epochs from 15 to 20 
February 2009 together with $R$-band flux for comparison.
No significant flux variation is found in all MIPS bands during those epochs, which include period D.

\begin{figure}[tbp]
  \centering
  \vspace{6mm}
 \includegraphics[width=7.7cm]{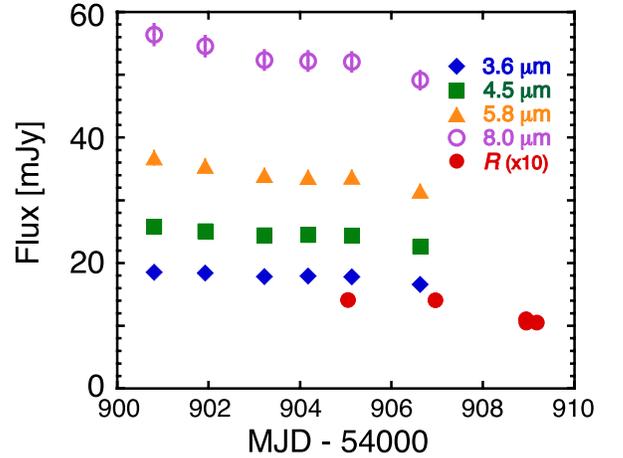}
  \caption{Light curve of 3C~279 at $3.6\,\mu$m, $4.5\,\mu$m, $5.8\,\mu$m and $8.0\,\mu$m measured by Spitzer-IRAC. 
  The error bars correspond to 3\% for all IRAC bands 
  as mentioned in the text in Section~\ref{sec:Spitzer}.
  Optical $R$-band data taken by the ground-based telescopes are also 
  plotted in red color for comparison.}
  \label{IRACLC}
 \end{figure}

The \Spitzer-IRAC calibration uncertainty is 3\,\% overall and has photometric 
repeatability of 1.5\,\%~\citep{rea05}.  We adopt the overall IRAC calibration 
uncertainty of 3\,\%, but note the following characteristics of our images.
In our IRAC frames, two standard comparison stars used in blazar monitoring 
were visible in the 3.6\,$\mu$m images (Star 1 and Star 
2)\footnote{see \url{http://www.lsw.uni-heidelberg.de/projects/extragalactic/charts/1253-055.html},
\citet{rai98}, \citet{vil97} and 
\url{http://quasar.colgate.edu/\~tbalonek/optical/3C279compstars.gif}}.
One comparison star, Star 2, was visible in the 4.5, 5.8, and 8\,$\mu$m images, 
located at the interstice of the chopping regions where the data are noisier than 
elsewhere.  The spacecraft orientation, and hence the chopping orientation, was 180 degrees 
different between 2008 July--August and 2009 March. The standard deviations in 
comparison Star 2's measurements in 2009 March at 3.6, 4.5, 5.8 and 8\,$\mu$m are
0.68\,mJy, 0.08\,mJy, 0.08\,mJy and 0.09\,mJy (5\,\%, 1\,\%, 2\,\% and 3\,\%), 
respectively. The high 3.6\,$\mu$m standard deviation was affected by a single 
high value on 11 March 2009, for which we found no obvious cause; excluding 
that value resulted in a standard deviation of 0.06\,mJy (0.5\,\%).
In contrast, the flux of 3C~279 shows a steady decrease of 10\,\%, 12\,\%, 14\,\% and 13\,\% 
at 3.6, 4.5, 5.8 and 8.0\,$\mu$m, respectively, during the 6 epochs from 2009 March 10--16 as shown in Figure~\ref{IRACLC}.

\subsection{Radio: CARMA}

Observations were obtained at mean frequencies of 92.5 and 227.5 GHz
using the Combined Array for Research in Millimeter-wave Astronomy
(CARMA; \citealt{boc06}). In all cases the nominal signal to noise  
ratio exceeded 400 and calibration
uncertainties dominated the errors. The source is bright enough to  
permit self-calibration on timescales of less than a minute
and so atmospheric decorrelation was not expected to affect our results
significantly even at the long baselines.  However, observations in
poor weather were not used due to the difficulty of reliably measuring
pointing offsets in these conditions.

Data calibration and analysis was done with the MIRIAD software
package \citep{sau95}. Flux densities were determined by first using  
phase self calibration with a short enough averaging interval to  
avoid any atmospheric phase de-correlation, then the flux density was  
determined from the vector average fringe amplitude at the position  
of 3C~279 over all baselines. For a strong point source such as  
3C~279 this provides very robust and unbiased amplitude estimate  
independent of the weather or the interferometer baselines.
We rely on regular system temperature measurements to provide flux  
calibration relative to the the fixed system sensitivity. The  
absolute flux calibration of CARMA observations is usually quoted as  
10--15\,\%. However, based on measurements made on the blazar 3C~454.3  
we estimated the relative flux calibration at each frequency to be  
within 5\,\% at 3 mm and 10\,\% at 1 mm.
The radio fluxes at 92.5 and 227.5\,GHz measured by CARMA correspond to
$F_{92.5\,{\rm GHz}} = 11.7 - 19.7$ Jy (14 data points) 
and $F_{227.5\,{\rm GHz}} = 6.3 - 9.2$ Jy (14 data points).
Figure~\ref{radioLC} includes the flux history of those radio data.

\subsection{Radio: OVRO 40\,m}

The Owens Valley Radio Observatory (OVRO) 40~m radio data were collected as part of an ongoing
long-term, fast-cadence $\gamma$-ray blazar monitoring campaign,
described in detail in \citet{Richards_2011}. Flux densities were
measured in a 3~GHz bandwidth centered on 15.0~GHz using dual,
off-axis 2\farcm 5 FWHM beams with 12\farcm 95 separation. Dicke
switching against a blank sky reference field to remove gain fluctuations and
atmospheric and ground contamination were
used.  Flux densities from this program are found to have a
minimum uncertainty of 4~mJy (mostly thermal) and a typical
uncertainty of 3\,\% for brighter sources.  During the period included
here, 3C~279 was observed as a pointing calibrator.  The flux density
scale was referred to the value for 3C~286 \citep[3.44~Jy
at 15~GHz;][]{Baars1977} with a scale uncertainty of about 5\%.
The radio flux at 15\,GHz measured by OVRO was ranging from 
11.1 to 18.0 Jy among 124 data points during the 2-year observations.
The light curve of the OVRO radio data is also plotted in Figure~\ref{radioLC}.

\section{Results of the multi-wavelength observations}
\label{section:results}


\begin{figure*}[htpb]
  \centering
 \includegraphics[width=11.5cm]{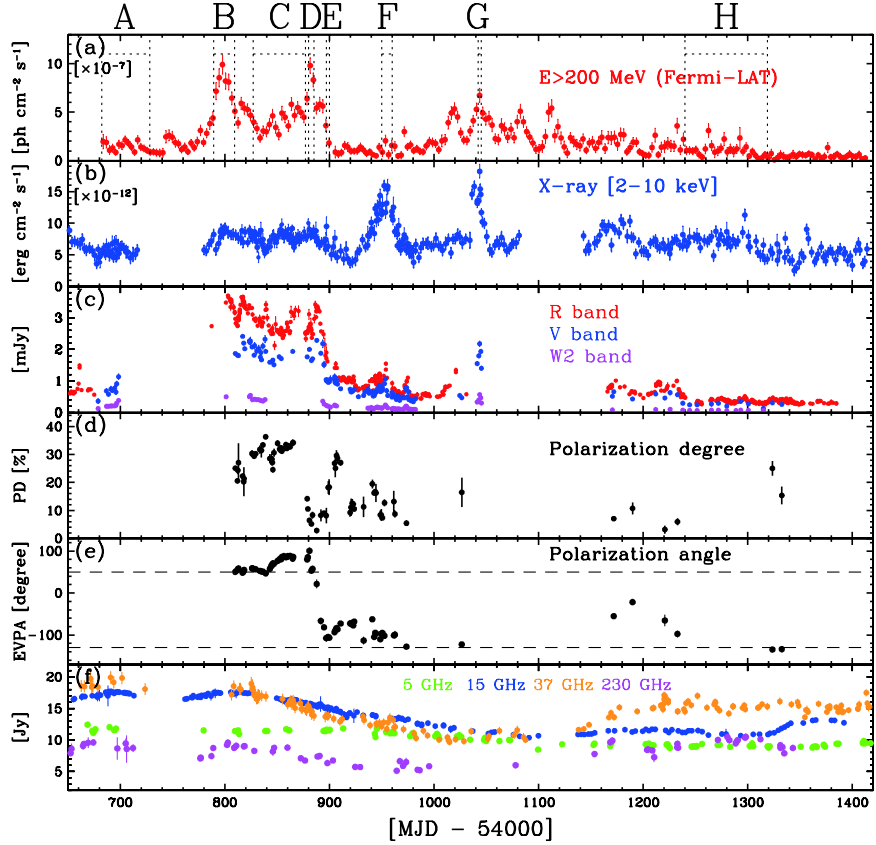}
\caption{Multi-band light curves of 3C~279 for 2 years from 2008 August to 2010 August.
  (a): Gamma-ray flux above 200 MeV averaged over 3 days. 
  (b): X-ray flux between 2 and 10 keV measured by \Swift-XRT 
  and \RXTE-PCA. (c): UV-optical fluxes in $R$-band 
  (red), $V$-band (blue) and $W2$-band (magenta). 
  (d): Polarization degree in the optical band.
  (e): Polarization angle in the optical band. The horizontal dashed lines refer 
  to the angle of 50\degree\ and $-130$\degree. 
  (e): Radio fluxes in 
  230 GHz band (magenta), 37 GHz band (orange), 15 GHz band (blue) and 5 GHz band (green). 
  All X-ray, UV and optical data are corrected for the Galactic absorption.
}
  \label{GXORLC}
 \end{figure*}

\subsection{Correlations of light curves in various bands}
\label{section:lightcurves}

The multi-band light curves of 3C~279 are presented in Figure~\ref{GXORLC}.
They include (a) $\gamma$-ray flux above 200 MeV (\Fermi-LAT), 
(b) 2-10 keV X-ray flux measured by \Swift-XRT and \RXTE-PCA, 
(c) optical-UV fluxes in $R$-band (GASP), $V$-band (\Swift-UVOT and Kanata) and $W2$-band (\Swift-UVOT),
(d,e) degree and angle of optical polarization (Kanata and KVA) and (f) radio 
fluxes in the 230, 37, 15 and 5 GHz bands (GASP, CARMA, and OVRO). We note that the X-ray 
fluxes determined by \Suzaku\ and \XMM\ are entirely consistent with those
plotted in Figure~\ref{GXORLC}. 
The extensive data set obtained in 
many bands for 3C~279 allows us to make general statements regarding the relative
flux variability in various spectral bands, and the relationship of the time
series to each other. The first such feature of the multi-band light curves
is a general - although not exact - trend where the IR through optical emission 
seems to be correlated with the $\gamma$-ray flux.
We calculated the Discrete Correlation Function (DCF; \citealt{DCF})
to quantify the correlation of the flux variations between the $\gamma$-ray and other bands,
and to determine whether we can measure any clear lag between the bands.

\begin{figure}[htbp]
  \centering
  \vspace{3mm}
 \includegraphics[height=7.4cm, angle=270, clip]{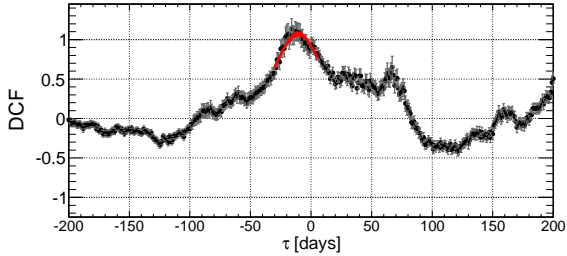}
  \caption{Discrete Correlation Function (DCF) derived for the $\gamma$-ray and optical $R$ bands.
Positive values of `$\tau$' correspond to flux variations in the $\gamma$-ray band lagging flux variation in the optical band.
The red curve represents 
a Gaussian fit to the data between $-30$ and 5 days. See the text for the fitting results.}
  \label{fig:DCFgo}
 \end{figure}

For the DCF calculations, we use the $\gamma$-ray fluxes averaged over an interval 
of 1 day as shown in the top panel of Figure~\ref{GammaLC_time}. 
The resulting
DCF between $\gamma$-ray and optical $R$-band fluxes 
is shown in Figure~\ref{fig:DCFgo}.
Positive values of `$\tau$' correspond to 
flux variations in the $\gamma$-ray band lagging flux variations in the other bands.
In the DCF between $\gamma$-ray and optical $R$-band fluxes, a peak can be seen 
close to zero lag. We fit the DCF data points in the range between $-30$ 
and 5 days using a Gaussian function of the form 
${\rm DCF}(\tau)= C_{\rm max}\times\exp[(\tau-\tau_0)^2/\sigma^2]$, where $C_{\rm max}$ 
is the peak value of the DCF, $\tau_0$ is the time at which the DCF 
peaks, and $\sigma$ is the Gaussian width of the DCF.
The fit yields a position of the peak at $\tau_0 = -10.7\pm0.7$ days, 
corresponding to a value of $C_{\rm max} = 1.07\pm0.03$ with 
a dispersion of $\sigma = 19.4\pm1.4$ days.  
The result implies that the optical emission is possibly delayed 
with respect to the $\gamma$-ray emission by about 10 days.

In the framework of the one-zone synchrotron + external-radiation Compton (ERC) 
models, the same electron population, of roughly the same energies, is responsible 
for the radiation in both the optical and $\gamma$-ray bands.
There, the 
observed lag can result from 
different profiles of the decreasing magnetic 
and radiation energy densities along the jet:  we show that idea quantitatively in Appendix A.  
As is shown there, a very steep drop of the external radiation energy density is required to explain the lag
in a conical jet with magnetic field $B' \propto 1/r$ where $r$ is the distance along the jet.
This condition can be relaxed in the scenario involving 
the re-confinement of a jet~\citep[e.g.,][]{key:DM88,key:KF97,key:NS09}.
In such a case, the magnetic field intensity can drop more slowly than $1/r$.
If the lag of the optical emission is confirmed, the application of the results in Appendix A to 
the $\sim 10$-day lag may imply the location of the active 
``blazar zone'' at distances of a few pc in agreement with those postulated to explain 
the optical polarization swing (Period D-E) in terms of a region containing an enhanced density of 
ultra-relativistic electrons propagating along a curved trajectory (Paper I).  
It is worth noting that similar $\gamma$-ray/optical lags have been reported 
during the outbursts of 3C~279 in early 1999~\cite{Har01vari}, of PKS~1502+106 in 2008 \citep{PKS1502}, of PKS~1510-089 
in early 2009 \citep{PKS1510, DAm11} and of AO~0235+164 in late 2008~\citep{Agudo0235, A0235}.  
On the other hand, no significant lags between 
$\gamma$-ray and optical signals have been detected in 3C~454.3 in late 2008 
\citep{Bonning09,Jor10}, in 3C~66A in 2008 October \citep{3C66A} and in OJ~287 
in 2009 October \citep{OJ287}.
Based on investigations of long-term light curves of 3C~454.3 during 2008--2010, 
\citet{rai11} have shown that the optical and $\gamma$-ray flux variations 
are not always simultaneous and have proposed a geometrical scenario 
to explain the change in the $\gamma$/optical flux ratio during the outburst peaks in 3C~454.3.
It is expected that the on-going multi-band 
monitoring of blazars will 
enable us to quantify such lags and find out how common they are.

 \begin{figure}[htbp]
  \centering
 \includegraphics[height=7.4cm, angle=270, clip]{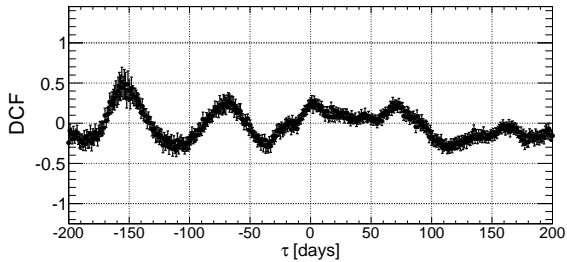}
  \caption{DCF between derived for the $\gamma$-ray and X-ray bands.
  Positive values of `$\tau$' correspond to flux variations in the $\gamma$-ray band lagging flux variation in the X-ray band.}
  \label{fig:DCFgx}
 \end{figure}

Different behavior is apparent in the radio flux, 
where the energies of radio-emitting electrons are very different from 
the energies of the electrons involved in producing the observed 
optical and $\gamma$-ray emission.
Variability appears to be much less rapid, 
and the excess variance ($F_{\rm var}$; see definition in Eq.~\ref{eq:vari}) 
in the radio regime is quite modest: 
for instance, $0.145\pm0.004$ at 37\,GHz, $0.165\pm0.001$ at 15\,GHz
and $0.104\pm0.001$ at 5\, GHz.
Those values are
significantly 
less than ones in the $\gamma$-ray or optical bands.
This suggests that the synchrotron emission from the
$\gamma$-ray emitting region is self-absorbed at these wavelengths.
The observed radiation is produced at much larger distances,
where the light-travel effects smear out the sharp,
rapid variability patterns observed in the 
optical and $\gamma$-ray bands.  

Perhaps the most surprising behavior - and difficult to 
explain in the context of simple, one-component, single-zone models - is the 
relationship of the X-ray light curve to those in the IR-optical 
or $\gamma$-ray bands. In Paper I, we reported that the X-ray time series 
exhibits a relatively rapid, symmetrical flare at $\sim$ MJD~54950 (Period F) with 
a duration of $\sim 20$ days, which is {\sl not} accompanied with 
any prominent IR/optical or $\gamma$-ray flares.  As we argued in Paper I,
the hard (rising in $\nu F_\nu$ representation) 
X-ray spectrum is unlikely to be the ``tail'' of the synchrotron emission, 
but instead, it is more likely to be produced by the low-energy end of 
the electron distribution radiating via inverse Compton process.

The continuing 
monitoring of the object in the X-ray band revealed another X-ray flare 
at $\sim$MJD~55040 {\bf (Period G)}, 
$\sim 90$ days after the first X-ray flare. The separation of the two 
X-ray flares is remarkably close to the temporal separation of the two 
$\gamma$-ray flares, with the two pairs delayed with 
respect to each other by $\sim 155$ days.  
Figure~\ref{fig:DCFgx} presents
the calculated DCF between $\gamma$-ray 
and X-ray fluxes, 
which shows a modest peak at $\sim-155$ day 
with a correlation coefficient of 0.6--0.7 and indicates no 
correlation between the the $\gamma$-ray and the X-ray bands with zero lag.
While confirming the physical connection of the two 
pairs would be very important, we cannot currently envision any situation 
where the two would be causally connected:  the $155$-day lag 
would imply the distance of the X-ray flare production 
$\sim 155 \Gamma_{\rm j}^2\;{\rm light\,days} \sim 50 (\Gamma_{\rm j}/20)^2$ pc and at such 
a distance should be accompanied by radio flares, which are not seen in our data.
In such a scenario, the X-ray flares should be significantly broadened compared 
to the $\gamma$-ray flares, however, we observe a similar temporal structure in both bands.
Furthermore, we note that 
there are some optical and
$\gamma$-ray peaks that might well be associated with the second X-ray flare.
Hence, it is possible that 
the two prominent $\gamma$-ray/optical flares (Period B and D), 
together with the subsequent two X-ray flares  (Period F and G),
form a sequence of 4 events
separated by a similar time intervals.
Those intervals, in turn, can be possibly determined by instabilities
in the jet launching region. Here, the different broad band spectra during
these events may result from small changes of parameters, such as
the jet direction, Lorentz factor, and/or location and geometry of 
the dissipation event. 

A weak (and sporadically almost absent) correlation between X-rays and
other spectral bands can also 
result from such processes which preferably contribute to radiation
in the X-ray band. They can be related to the following three
mechanisms/scenarios:
\begin{enumerate}
\item Bulk-Compton process. This involves Compton-scattering of ambient 
optical/UV light by the {\sl cold} (non-relativistic) electrons in the jet.
This mechanism is most efficient close to the accreting black hole where the
processes responsible for the variability of X-rays may operate independently 
of those at larger distances and producing there variable
non-thermal radiation~\citep{beg87}. A drawback of this scenario can be that 
the bulk-Compton spectrum is predicted to have a similar shape as the 
spectrum of the external radiation field~\citep{A0235},
which significantly differs from what we observe in the X-ray band.  
\item
Inefficient electron acceleration. 
Acceleration of the relativistic electrons
at proton-mediated shocks is 
 likely to proceed in two steps: 
in the first one 
low-energy electrons may be pre-accelerated via, for 
example, some collective processes involving protons;
in the second step, they may participate in 
the first-order Fermi acceleration process.
If under some conditions the electron-proton coupling is inefficient, 
the fraction of electrons reaching the Fermi phase of acceleration 
will be small. In this case the X-rays, originating from lower energy
electrons, are produced efficiently, while the $\gamma$-rays and optical 
radiation which involve more relativistic electrons - are not.  
\item The X-rays can be also contributed
by hadronic processes, specifically by the pair cascades powered by protons
losing their energy in the photo-mesonic process~\citep{Man92}.
For this process to be efficient, it requires extreme conditions~\citep{Sik09, Sik11},
however, operating in the very compact central region, at distances less than
few hundred gravitational radii, it may occasionally dominate in the X-ray band.
\end{enumerate}

\subsection{Broad-band Spectral Energy Distribution}
\label{section:SED}

\begin{figure*}[htbp]
  \centering
 \plotone{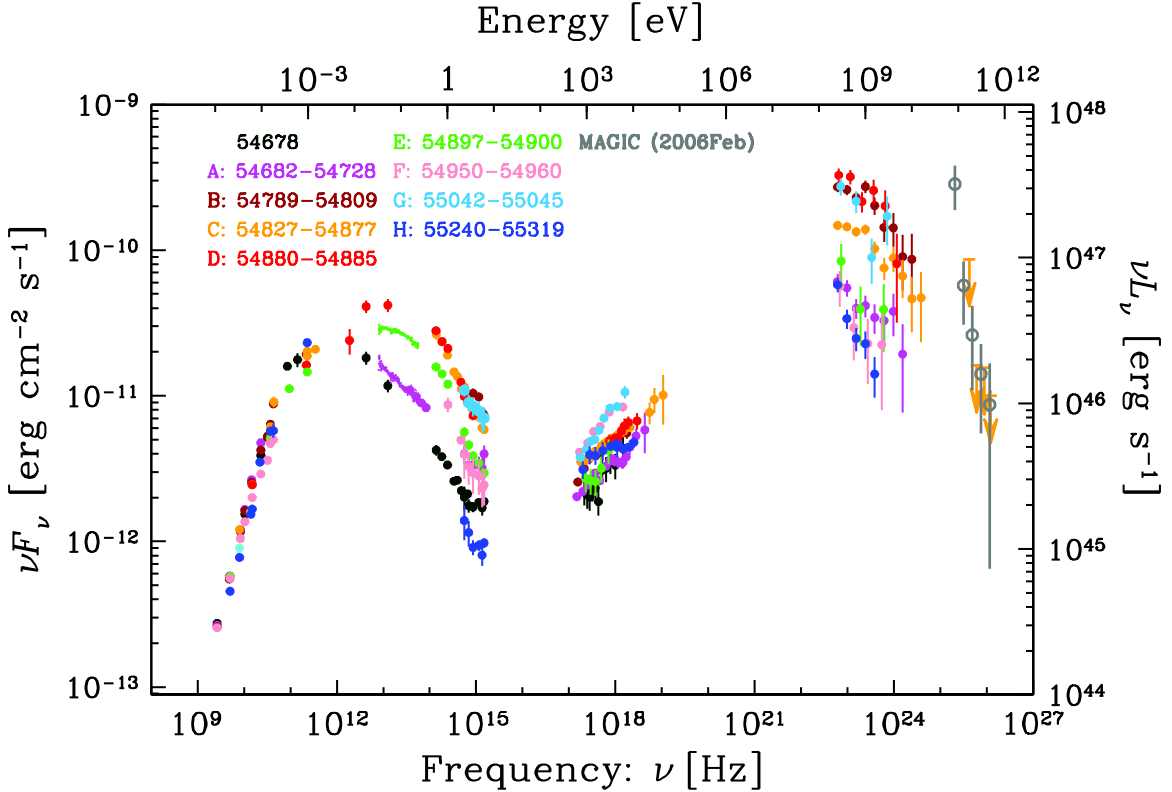}
  \caption{Time-resolved broad-band spectral energy distributions of 3C~279
  measured in Period A--H (as defined in Table~\ref{GammaSPfit})
  and on 2008 July 31 (MJD 54678), 
  covered by our observational campaigns in 2008--2010.
  X-ray, UV-optical-nearIR data are corrected for the Galactic absorption.
  5-digit numbers in the panel indicate MJD of the periods.
  For comparison, the gray open circles in the very-high-energy $\gamma$-ray band
  represent measured spectral points by MAGIC in 2006 February~\citep{MAGIC}.
  }
  \label{allSEDone}
 \end{figure*}

\begin{figure*}[htbp]
  \centering
 \plotone{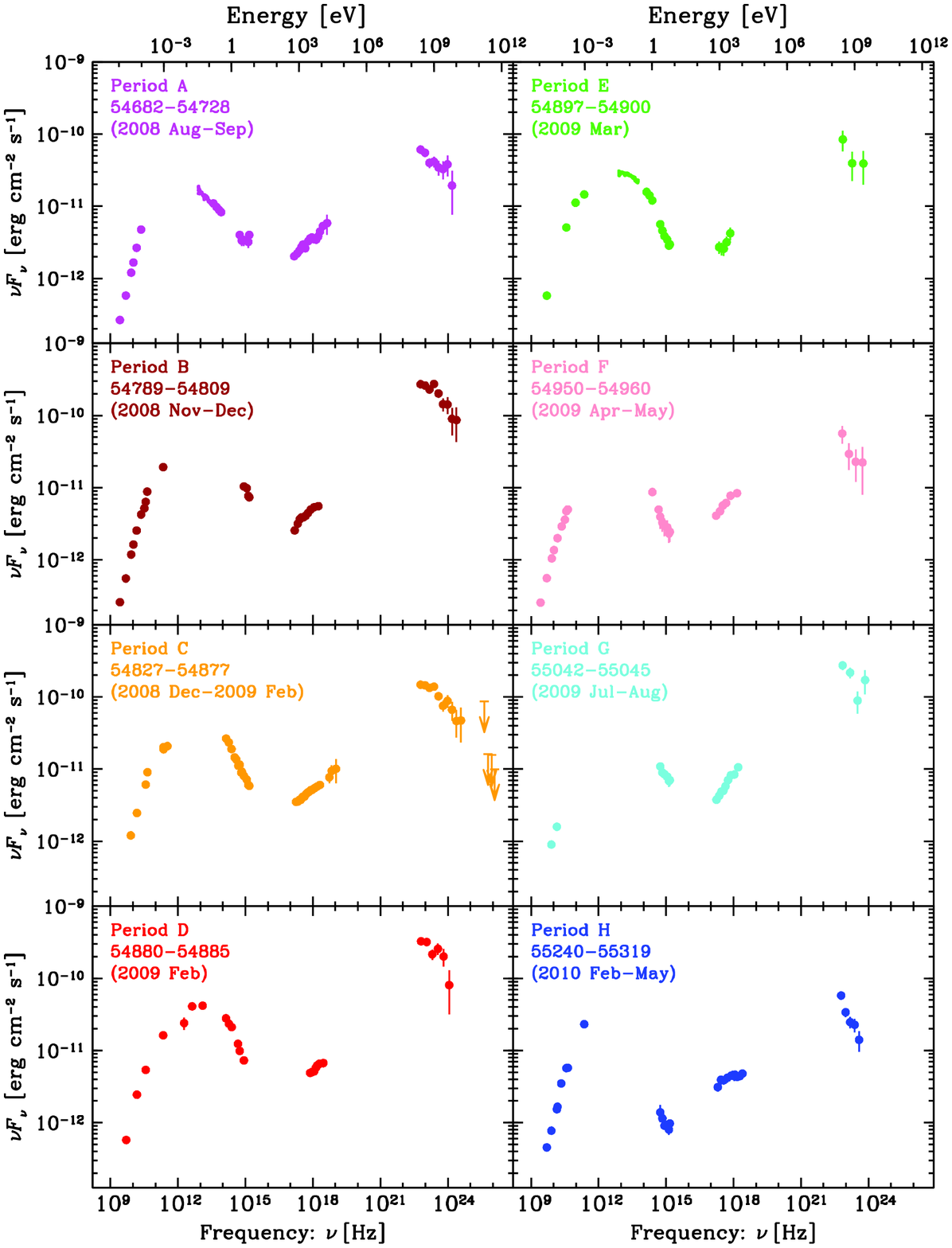}
  \caption{Time-resolved broad-band spectral energy distributions of 3C 279 in each period (A--H), covered by the campaigns.
  The data points are the same as ones in Fig.~\ref{allSEDone}, but are plotted in a separate panel for each period.
  5-digit numbers in each panel indicate MJD of the observing period 
  of each broad-band spectrum.
  }
  \label{allSED}
 \end{figure*}

Figures~\ref{allSEDone} and~\ref{allSED} show broad-band SEDs of 3C~279 in all 
periods as defined in Table~\ref{GammaSPfit}. In addition, we also 
extracted a SED using data taken on 2008 July 31 (MJD~54678), which has a good energy 
coverage of the synchrotron emission component including \Spitzer\ and GROND data,
although the $\gamma$-ray data by \Fermi-LAT are not available at that time because this was before the 
beginning of normal, all-sky science observations with \Fermi-LAT.
Both SEDs for Period D (2009 February: corresponding to the brightest $\gamma$-ray flare coincident 
with the optical polarization swing) and Period F (2009 April: corresponding to the first 
isolated X-ray flare) have already been partially reported in Figure 2 of Paper I.
New \Spitzer-MIPS data points are included in the SED for Period D in this paper.
In Period C, there are observations by MAGIC, which provide upper limits above 
100~GeV~\citep{MAGICUL}.  For comparison, we also include very-high-energy 
$\gamma$-ray fluxes detected with the MAGIC telescope in 2006 February as gray 
points~\citep{MAGIC} in Figure~\ref{allSEDone}.

   
This is the richest set of time-resolved spectra ever collected for this source.
The spectral coverage of the synchrotron bump is unprecedented, allowing us 
not only to constrain the parameters of the emission models, but also to study 
their time evolution.  
As we discussed in Section~\ref{sec:gamma_sp}, the shape 
of the $\gamma$-ray spectrum deviate from a simple power law, in similarity to 
other FSRQ blazars. Strong variability, over one order of 
magnitude, is evident in nearIR/optical/UV and $\gamma$-ray bands.
This contrasts with the moderate variability in the radio/mm and X-ray bands.

Particularly interesting is the behavior of this source in the mid-IR band, around $\sim 10^{13}$ Hz, 
where significant spectral variability is observed.  In the low state 
in Period~A the mid-IR spectrum is relatively soft and can be extended with a 
power law shape to the optical/UV band. In this case, the synchrotron component 
peaks in the mm/sub-mm band ($\sim 10^{11}-10^{12}$ Hz).  However, in the high state in Periods~D and E, the 
mid-IR spectrum is much harder and shows a significant curvature. In Period D,
there is a clear 
spectral break at $\sim 3\times 10^{12}$\,Hz ($\sim 100\,{\rm \mu m}$).  The spectral index between the
$70\,{\rm \mu m}$ ($\sim4.3\times10^{12}$\,Hz) and $160\,{\rm \mu m}$ ($\sim 1.9\times10^{12}$\,Hz) points is $\alpha_{70-160}=0.35\pm0.23$,
taking into account systematic errors described in Section~\ref{sec:Spitzer}.
The synchrotron peak is located in the mid-IR band, at a frequency one order 
of magnitude higher than in the low state. This indicates that there are two 
independent synchrotron bumps, possibly produced at different locations. 
The mid-IR-peaking 
component, seen only in the IR/optical/UV flaring state, is characterized by a strong and 
rapid variability. The mm/sub-mm peaking component is more persistent and 
dominates when the source is in the low state. 
The complex shape of the SED between 
the mm band and the $70\;{\rm \mu m}$ point in Period D requires a coexistence of 
these two components.  A similar scenario of multiple synchrotron components 
was investigated in the case of 3C~454.3 by \citet{Ogl10}.

In the X-ray band, despite the smaller variability amplitude, we observe 
some spectral changes.  In particular, in Periods F and G, which represent 
the two isolated X-ray flares, the spectrum is very similar and harder than on average. 
Figure \ref{allSEDone} shows that these flares are not energetically important.  
If we extrapolate the X-ray spectra with power laws to the $\gamma$-ray band, 
we under-predict the observed $\gamma$-ray flux in Periods B, C, D and H.  
Periods B, C, D coincide with the high-activity $\gamma$-ray state.  
This indicates that the X-ray flux cannot originate 
from the same emission component
as the $\gamma$-ray flux, at least in the flaring state.  
Because the $\gamma$ rays are correlated with the optical flux 
but not with the X-ray flux, the $\gamma$ rays can be related to the mid-IR-peaking 
synchrotron bump while the X-rays may correspond to the mm/sub-mm peaking 
synchrotron bump.  We explore this possibility when modeling the SEDs 
at Periods A and D in Section \ref{section:modeling:ercbel}.

\section{Modeling the broad-band emission}
\label{section:modeling}
 
We have fitted selected SEDs with one-zone leptonic models 
described in~\citet{Mod03}, 
including synchrotron emission and self-absorption, 
Comptonization of the local synchrotron radiation [SSC component] 
and external photons [ERC component], but also including the opacity
due to internal pair-production. The external radiation includes 
broad emission lines (BEL) and infrared dust emission (IR).
Their energy densities in the jet co-moving frame as functions of the distance 
$r$ along the jet are approximated by the formulae:
\begin{equation}
u_{\rm BEL}'(r)= \frac{\xi_{\rm BEL}\Gamma_{\rm j}^2L_{\rm D}}{3\pi r_{\rm BEL}^2 c [1+(r/r_{\rm BEL})^{\beta_{BEL}}]} \\
\end{equation}
\begin{equation}
u_{\rm IR}'(r)=\frac{\xi_{\rm IR}\Gamma_{\rm j}^2L_{\rm D}}{3\pi r_{\rm IR}^2c[1+(r/r_{\rm IR})^{\beta_{IR}}]}
\end{equation}
where $\xi_{\rm BEL}=0.1$ and $\xi_{\rm IR}=0.1$ are the 
fractions of the disk luminosity  
$L_{\rm D} \simeq 2\times 10^{45}\;{\rm erg\,s^{-1}}$ 
reprocessed into emission lines and into hot dust radiation, 
respectively, $r_{\rm BEL} = 0.1 (L_{D,46})^{1/2}$\,pc 
and $r_{\rm IR} = 2.5 (L_{\rm D,46})^{1/2}$\,pc 
[$L_{\rm D,46} \equiv L_{\rm D}/10^{46}$]
are the characteristic 
distances where such reprocessing takes place,
and $\Gamma_{\rm j}$ is the jet Lorentz factor.
The external radiation fields are approximated
in the jet co-moving frame by 
Maxwellian spectra peaked at photon energies 
$E_{\rm BEL}'\sim 10\;{\rm eV}\times\Gamma_{\rm j}$ and 
$E_{\rm IR}'\sim 0.3\;{\rm eV}\times\Gamma_{\rm j}$.
While the radiation density profile in the frame external to the jet 
should satisfy $\beta_{\rm BEL(IR)}\le 2$, it is not applied to the 
profile in the jet co-moving frame.  This is because the transformation of 
radiation density depends on the angular distribution of external radiation, 
with radiation arriving at small incidence angles to the jet velocity vector 
being actually deboosted. This can result in a steeper profile of the radiation 
density in the jet co-moving frame.  Here, we adopt  $\beta_{\rm BEL} = 3$~\citep{Sik09} 
and $\beta_{\rm IR} = 4$ (see Section \ref{section:modeling:ercir}).
We assume a conical jet geometry;
the magnetic field, assumed to be dominated by the 
toroidal component, is taken to decline with distance $r$ as 
$B'\propto 1/r$. Electrons are injected with a double-broken 
power law distribution $Q(\gamma)\propto\gamma^{-p}$ with $p=p_1$ 
for $\gamma<\gamma_{\rm br1}$, $p=p_2$ for 
$\gamma_{\rm br1}<\gamma<\gamma_{\rm br2}$ and $p=p_3$ 
for $\gamma>\gamma_{\rm br2}$. 
Their evolution, including 
injection at a constant rate as well as radiative and adiabatic cooling, 
is followed over a distance $\Delta r=r/2$, where $r$ 
is the position at which the injection ends. The emission is 
integrated over spherical thin shells within a conical 
region of opening angle $\theta_{\rm j}=1/\Gamma_{\rm j}$.
The observer is located within the jet opening cone, i.e. 
$\theta_{\rm obs}\lesssim\theta_{\rm j}$.

We begin by modeling the SED in Period D, which is the highest 
$\gamma$-ray state reached by the source during our observational campaigns.
In Paper I, we showed that the flare event was accompanied by an 
optical polarization swing and proposed two interpretations 
of this event.  The first one involved a cloud containing 
ultra-relativistic particles propagating along a curved trajectory.
The duration of the polarization swing constrains the location 
of the cloud to be at a few parsecs from the central 
supermassive black hole, in the region where external 
radiation is dominated by the infrared dust emission.  
In Section \ref{section:modeling:ercir}, we present an ERC-IR model 
describing the SED in Period D and a physically related model 
of the SED in Period E.  
The second interpretation of the polarization 
swing involved the jet precession,
which allowed arbitrary location 
of the emitting region, including the broad-line region.  
In Section \ref{section:modeling:ercbel}, we present an ERC-BEL 
model of the SED in Period D.  We show that in this scenario the 
far-IR break arises due to synchrotron self-absorption.
We also show an ERC-IR model of the SED in Period A, which 
can explain the mm/far-IR and X-ray emission, as well as 
the low-state optical and $\gamma$-ray flux levels.

We assume the scenario where the X-ray emission is unrelated to the flaring 
component, 
since it showed little variability during the correlated 
$\gamma$-ray/optical flares.  Our one-zone models of the 
flaring states are fitted only to the IR/optical/UV and $\gamma$-ray data, 
treating the simultaneous X-ray spectrum as only an upper limit 
to the SSC component and the ERC component from the low-energy electrons.
The large $\gamma$-ray/X-ray luminosity ratio forces us to adopt 
a very hard electron energy distribution at low energies ($p=1$), 
which can be alternatively obtained by imposing a minimum electron 
Lorentz factor $\gamma_{\rm min}\gg 1$.

\begin{figure}[bp]
\centering
     \plotone{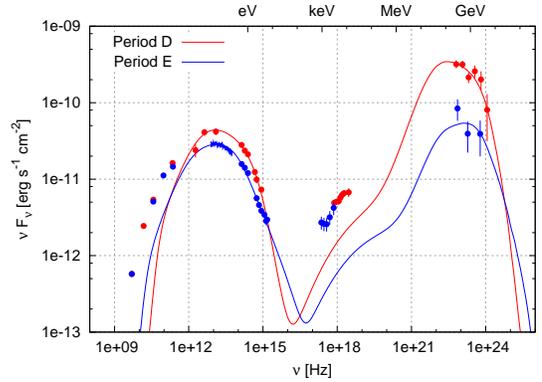}
   \caption{Emission models D1 (red line) and E1 (blue line) 
    fitted to the spectral states at Periods D and E, 
    respectively.
    Periods D and E correspond to the first 5 days and the last 3 days of the $\gamma$-ray flaring event 
    accompanied by an optical polarization change, respectively.
    Those models adopt our `propagation scenario', where external radiation is dominated by infrared dust emission. 
     We assume the X-ray emission is not related to the flaring component, 
    and consider the X-ray fluxes as only upper limits to the SSC and the ERC components during the flaring event.
    See the text in Section~\ref{section:modeling} and \ref{section:modeling:ercir} for details of the models and Table~\ref{tab_models} for model parameters.}
   \label{fig_models1}
 \end{figure}
 
 \begin{figure}[bp]
\centering
   \plotone{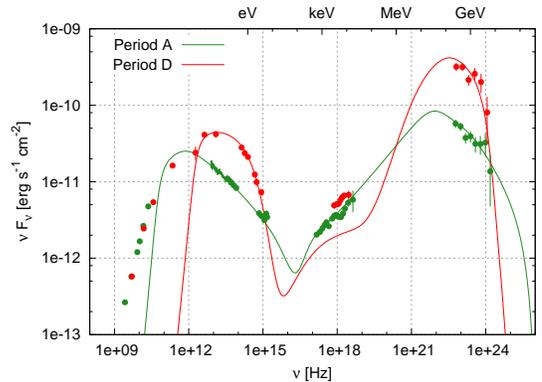}
   \caption{Emission Models A2 (green line) and D2 (red line) 
    fitted to the spectral states at Periods A and D, respectively. 
    Period A represents a quiescent state, and Period D corresponds to the $\gamma$-ray flaring event accompanied by an optical polarization change.
    Those models adopt our `jet precession scenario', which assumes the $\gamma$-ray flaring event (Period D) occurs within the broad-line region
    while the low-steady emission component (Period A) is generated outside the broad-line region.
    See the text in Section~\ref{section:modeling} and \ref{section:modeling:ercbel} for details of the models and Table~\ref{tab_models} for model parameters.}
   \label{fig_models2}
 \end{figure}

\subsection{Propagation scenario for the emitting region}
\label{section:modeling:ercir}

An intrinsically spherically symmetric emitting region 
is expected to produce the observed electric polarization 
vector aligned with the projected velocity of the emitting 
region. \citet{Nal10} presented a simple model of its trajectory 
to explain the event of simultaneous 
smooth variations of the polarization degree and angle during 
the polarization swing which has been reported in Paper I.
This model adopts a constant jet Lorentz 
factor $\Gamma_{\rm j}=15$ and can be used to predict the 
viewing angles for a given observation time. For Period D 
we estimate $\theta_{\rm obs,D}\sim 1.5^\circ$, while for 
Period E: $\theta_{\rm obs,E}\sim 2.4^\circ$. Between 
Periods D and E ($\Delta t\sim 15\;{\rm days}$) the emitting region 
propagates over a distance 
$\Delta r\sim\Gamma_{\rm j}^2c(\Delta t)\sim 2.8\;{\rm pc}$.

In Figure \ref{fig_models1} we show Model D1 fitted to the 
SED in Period D at $r=r_{\rm IR}$ and Model E1 fitted to 
the SED in Period E at $r=r_{\rm IR}+\Delta r$.  
Model parameters are listed in Table \ref{tab_models}.
Both models use the magnetic field scaled to the 
same value at the distance of 1 pc.  
In order to explain the difference in the luminosity ratio 
of the ERC component and the synchrotron component, which 
decreased by factor $\sim 4$ between Model D1 and Model E1, 
we assume a distribution of the co-moving IR radiation energy 
density dropping steeply with distance, 
adopting $\beta_{\rm IR}\sim 4$.
This corresponds to a strongly stratified torus structure, with 
a significant concentration of hot dust very close to the 
sublimation radius \citep[see, e.g.,][]{Mor11}.
We should note that,
although the relatively soft $\gamma$-ray spectrum was
observed at Period E ($\Gamma = 2.64\pm0.32$, see Table \ref{GammaSPfit}), 
the peak of the ERC-IR component in the Model~E1
falls at $\sim 800\;{\rm MeV}$, in the \Fermi-LAT band.

The far-IR spectral break in Period D requires a sharp 
break in the electron distribution function at $\gamma_{\rm br1}=800$.
As the cooling break is expected at $\gamma_{\rm c} \sim 3m_e c^2 / (2\sigma_T R u'_{ext})\sim 660$
(where $R$ is a radius of source emission region and $u'_{ext}$ 
is the energy density of the external radiation in the jet co-moving frame),
$\gamma_{\rm br1}$ is located just within the fast-cooling regime.
The electron distribution in the fast-cooling regime cannot 
be harder than $p=2$, hence the resulting synchrotron spectral 
index $\alpha=(p-1)/2$ should be larger than 0.5.
In fact, the mean value of the observed spectral 
index between 70\,$\mu$m and 160\,$\mu$m is smaller than 0.5 ($\alpha_{70-160} = 0.35\pm0.23$), 
which cannot be explained if the electron cooling is efficient.
However, the uncertainty of the measurement does not allow us to reject this scenario.

 \begin{deluxetable}{c|cc|cc}
   \tablecaption{Parameters of emission models.\label{tab_models}}
   \centering
   \startdata
     Model & D1 & E1 & D2 & A2 \\
     \hline
     ext. rad.\tablenotemark{a}& IR & IR & BEL & IR \\
     $r$ [pc] & 1.1 & 3.9 & 0.045 & 3.9 \\
     $R$ [pc]\tablenotemark{b} & 0.07 & 0.26 & 0.0023 & 0.19 \\
     $\Gamma_{\rm j}$ & 15 & 15 & 20 & 20 \\
     $\theta_{\rm j}\;[^\circ]$ & 3.8 & 3.8 & 2.9 & 2.9 \\
     $\theta_{\rm obs}\;[^\circ]$ & 1.5 & 2.4 & 1.7 & 1.7 \\
     $B'_{\rm 1 pc}$ [G]\tablenotemark{c} & 0.14 & 0.14 & 0.15 & 0.15 \\
     $u'_{\rm ext}$ [$10^{-4}$ erg cm$^{-3}$]\tablenotemark{d} & 78 & 0.97 & $8\times10^4$ & 1.8 \\
     $\gamma_{\rm br1}$ & 800 & 800 & 170 & 440\\
     $\gamma_{\rm br2}$ & 6500 & 5000 & 1000 & \nodata \\
     $p_1$ & 1 & 1 & 1 & 2.2 \\
     $p_2$ & 2.5 & 2.6 & 2.4 & 3.4 \\
     $p_3$ & 6 & 4.2 & 7 & \nodata 
  \enddata
  \tablenotetext{a}{A dominant component of the external radiation. IR: infrared dust emission, BEL: broad emission lines.}
  \tablenotetext{b}{Radius of source emission region.}
  \tablenotetext{c}{Magnetic field intensity at the distance of 1\,pc.}
  \tablenotetext{d}{Energy density of the dominant component of the external radiation in the jet co-moving frame at the given distance $r$.}
\end{deluxetable}

\subsection{The jet precession scenario:  two synchrotron components}
\label{section:modeling:ercbel}

Alternatively, 
if the jet precession can cause the observed $\gamma$-ray flare event with the polarization swing,
the $\gamma$-ray/optical emission can be generated much closer to the central black hole, 
even within the broad line region (see also in Paper I).
Therefore, we also attempted to model Period~D
placing the emitting region at $r_{\rm BEL}$.
For $\Gamma_{\rm j}=15$, with model parameters fitted using the 
synchrotron and ERC components, the X-ray flux is overproduced 
by the SSC process. To alleviate this problem, we increased the jet Lorentz 
factor to $\Gamma_{\rm j}=20$. In Figure \ref{fig_models2},  
we show Model D2 with parameters listed in Table \ref{tab_models}. 
The magnetic field strength scaled to the distance of $1\;{\rm pc}$ 
is almost the same as the value in Model D1.  
Because of a smaller size of emission region and 
higher energy density 
of the locally produced synchrotron radiation,
the synchrotron self-absorption is able to produce
a spectral cut-off at a higher frequency of $\sim 3\times 10^{12}\;{\rm Hz}$ 
($\sim 100\;{\rm \mu m}$), consistent with the far-IR break.  
This interpretation 
has an advantage that it also could explain the observed hard spectral 
index between 70\,$\mu$m and 160\,$\mu$m, even smaller than 0.5,  independently of
details of the electron energy distribution.
 
The low-energy synchrotron component, dominating the mm/sub-mm band, 
must be produced in a much larger region, placing it far outside the 
broad-line region.  In Figure \ref{fig_models2}, we present Model A2, 
fitted to the SED at Period A. We kept the Lorentz factor and 
the magnetic field consistent with Model D2, but we set the 
source at the distance $\sim 4\;{\rm pc}$, the same as in Model E1.
This low state model of Period~A can reproduce both observed X-ray and $\gamma$-ray spectra
by a single broken power law electron distribution. 
The $\gamma$-ray spectral index is consistent with 
the IR/optical/UV spectral index.  The synchrotron self-absorption 
is effective at $\sim 10^{11}\;{\rm Hz}$ and the spectral peak is 
located in the mm/sub-mm band.

Those results suggest the existence of two synchrotron components:
one peaking in the mm/sub-mm band and the other peaking in the mid-IR band.
The component with the peak in the mid-IR band is more variable, and can be 
produced at shorter distances, within the broad-line region, where 
the far-IR break can be explained by the synchrotron self-absorption.

\section{Conclusions}
\label{section:conclusions}

This paper reports details of the multi-band campaigns on the well-known 
blazar 3C~279 during the first two years of the 
{\it Fermi} mission between 2008 and 2010. 
Some key results were already presented in \citet[Paper I]{paper1}.
Most important of them was the coincidence of a dramatic $\gamma$-ray/optical 
flare with a change in the optical polarization, which we interpreted 
as the result of a compact emitting region: either propagating along a curved 
relativistic jet or located at a constant distance in a precessing jet.  In addition, 
we reported on an ``isolated'' X-ray flare, an event without a clear counterpart 
in other bands,  and taking place a few months after the $\gamma$-ray/optical flare. 
Here, we extended the observation epoch until 2010 August
yielding the best coverage of time-resolved SEDs 
ever collected for 3C~279 from radio through high-energy $\gamma$-ray bands.
Based on those data we arrived at several new conclusions 
about the structure and emission models of the relativistic jet in the quasar:

\begin{itemize}

\item In the high-energy $\gamma$-ray band measured by \Fermi-LAT, 
the source exhibited two prominent flares reaching as high as $\sim 3\times10^{-6}$\,\phcms\ above 100\,MeV in the first year while it was in a relatively quiescent state in the second year.
No significant correlation between flux and photon index has been measured in similarity to other LAT blazars.
The 2-year averaged $\gamma$-ray spectral shape above 200~MeV clearly deviates from a single power law.
The broken-power law model returns a break energy within a few GeV range, which does not appear to vary with the source flux. 
Such behavior is similar to that observed in other bright FSRQs.

\item The superb temporal coverage allowed us to measure in detail the cross
correlation of the $\gamma$-ray and optical fluxes. The optical signals appear to be
delayed with respect to the $\gamma$-ray signals by $\sim 10$ days.
Such a lag can be explained in terms of the simple synchrotron
and inverse-Compton model, in the scenario where a cloud containing ultra-relativistic 
electrons propagates 
down the jet through the regions where the ratio of the external radiation 
energy to the magnetic energy densities decrease with distance. 
We have verified this idea qualitatively (see Appendix A), 
but it still needs specific numerical modeling to be confirmed quantitatively.

\item X-ray observations reveal a pair of pronounced flares separated by $\sim 90$ 
days.  Those are not contemporaneous with a pair of bright $\gamma$-ray/optical 
flares - also separated by $\sim 90$ days - but instead, are delayed with respect 
to the $\gamma$-ray/optical flares by about 155-days.  Because of such a long 
delay, it seems implausible that these events are causally related. Instead, 
the possible scenarios of the X-ray flares may involve changes of the source parameters 
such as the jet direction, Lorentz factor, and/or location of the dissipation event, 
or may require more 'exotic' solutions, for instance: bulk-Compton process; 
inefficient electron acceleration above a given energy; hadronic processes.
At this stage we cannot discriminate among any of those scenarios.

\item The spectral coverage of the infrared band with {\it Spitzer} 
enabled us to probe the detailed structure of the low-energy spectral bump, attributed 
to the synchrotron radiation. Significant spectral variability, 
with soft/power-law spectra in the low state and hard/curved spectra in the high state,
as well as the detection of a sharp far-IR spectral break in the high state,
strongly suggest the existence of two synchrotron components:
one peaking in the mm/sub-mm band and the other peaking in the mid-IR band.
The component with a peak at the mid-IR band can be responsible for emission during
$\gamma$-ray flaring states.

\item 
We have applied our leptonic emission model for the SEDs during the $\gamma$-ray flaring state with a polarization change.
Adopting the interpretation of the polarization swing involving the
propagation of the emitting region - that suggested in Paper I - 
we can explain the evolution of the broad-band SEDs from Periods D to E during the $\gamma$-ray flaring event
by a shift of the position of the emitting region 
and a change of the viewing angle that are consistent with its trajectory.
We used the same distribution of magnetic fields and only 
slightly changed electron spectra, but required a rather steep stratification 
of the external radiation density in the form of thermal emission from the dusty torus.
In this case, the far-IR spectral break requires a break in the electron distribution.
The observed \Spitzer-MIPS spectral index $\alpha_{70-160}=0.35\pm0.23$ is marginally 
consistent with the synchrotron emission in the fast-cooling regime.

\item We also discussed the model in which the $\gamma$-ray flare is generated within the broad emission line region
at sub-pc scale from the central black hole according to the jet precession scenario.
This model explains the mid-IR break during the flaring state of Period D by synchrotron self-absorption.
Here, we also discussed the low-state SED in Period A where the mm/sub-mm-band-peaking synchrotron component can be dominant.
The model shows the related ERC component can explain the steady X-ray emission.

\end{itemize}

\acknowledgements
The \textit{Fermi}-LAT Collaboration acknowledges generous ongoing support
from a number of agencies and institutes that have supported both the
development and the operation of the LAT as well as scientific data analysis.
These include the National Aeronautics and Space Administration and the
Department of Energy in the United States, the Commissariat \`a l'Energie Atomique
and the Centre National de la Recherche Scientifique / Institut National de Physique
Nucl\'eaire et de Physique des Particules in France, the Agenzia Spaziale Italiana
and the Istituto Nazionale di Fisica Nucleare in Italy, the Ministry of Education,
Culture, Sports, Science and Technology (MEXT), High Energy Accelerator Research
Organization (KEK) and Japan Aerospace Exploration Agency (JAXA) in Japan, and
the K.~A.~Wallenberg Foundation, the Swedish Research Council and the
Swedish National Space Board in Sweden.
Additional support for science analysis during the operations phase is gratefully
acknowledged from the Istituto Nazionale di Astrofisica in Italy and the Centre National d'\'Etudes Spatiales in France.

The Submillimeter Array is a joint project between the Smithsonian 
Astrophysical Observatory and the Academia Sinica Institute of Astronomy and 
Astrophysics and is funded by the Smithsonian Institution and the Academia 
Sinica.
St.Petersburg University team acknowledges support from Russian RFBR foundation via grant 09-02-00092.
AZT-24 observations are made within an agreement between Pulkovo,
Rome and Teramo observatories.
This paper is partly based on observations carried out 
at the German-Spanish Calar Alto Observatory, 
which is jointly operated by the MPIA and the IAA-CSIC.
Acquisition of the MAPCAT data is supported in part 
by MICIIN (Spain) grant and AYA2010-14844, 
and by CEIC (Andaluc\'{i}a) grant P09-FQM-4784.
The Mets\"ahovi team acknowledges the support from the Academy of Finland
to our observing projects (numbers 212656, 210338, 121148, and others).
The Medicina and Noto telescopes are operated by
INAF - Istituto di Radioastronomia.
The research at Boston U. was supported by NASA Fermi GI grants NNX08AV65G, NNX08AV61G, 
NNX09AT99G, and NNX09AU10G, and NSF grant AST-0907893. 
The PRISM camera at Lowell Observatory was developed by K.\ Janes et al.\ at BU and Lowell Observatory, 
with funding from the NSF, BU, and Lowell Observatory. The Liverpool Telescope is operated 
on the island of La Palma by Liverpool John Moores University in the
Spanish Observatorio del Roque de los Muchachos of the Instituto de Astrofisica de Canarias, with funding
from the UK Science and Technology Facilities Council.
The Abastumani Observatory team acknowledges financial support 
by the Georgian National Science Foundation 
through grant GNSF/ST09/521 4-320.
The research at U.~Michigan has been funded by a series of grants from NASA and from the NSF. Specific grant
numbers are NASA grants NNX09AU16G, NNX10AP16G, NNX11AO13G, and NSF grant AST-0607523.
Funding for the operation of UMRAO was provided by the U.~Michigan.

M.~H. is supported by the Research Fellowships of the Japan Society for the Promotion of Science for Young Scientists.
This work was partially supported by the Polish MNiSW grants N~N203 301635 and N~N203 386337, 
the Polish ASTRONET grant 621/E-78/SN-0068/2007, and the Polish NCN grant DEC-2011/01/B/ST9/04845.

\appendix
\section{Possible explanation of a lag between the $\gamma$-ray and
optical flares}

As we discussed in Section~\ref{section:lightcurves}, the multi-band time
series imply that during the flaring activity detected in 3C~279 the
optical emission appears delayed with respect to the $\gamma$-ray
emission. In the context of radiation models adopted here
(Section~\ref{section:modeling}), the same electron population produces
optical synchrotron photons and also inverse-Compton $\gamma$-rays
\emph{in the fast-cooling regime}. A lag between the optical and
$\gamma$-ray flares may therefore result from different profiles of
decrease of the magnetic energy density $u_{\rm B}'(r)$ and the external
(target) radiation energy density $u_{\rm ext}'(r)$ with the distance $r$
along the jet, convolved with a non-monotonic profile of the electron
injection rate within the outflow.

In the fast-cooling (FC) regime, the power injected into the relativistic
particles $P_{\rm e,\,inj}$ is immediately radiated away and determines
the total broad-band luminosity produced by the cooled electrons $L_{\rm
tot,\,FC}$. Assuming a strong inverse-Compton dominance, i.e., the
observed inverse-Compton peak ($\gamma$-ray) luminosity $\simeq
L_{\gamma}$ being much larger than the observed synchrotron peak (optical)
luminosity $L_{\rm opt}$, one has
\begin{equation}
P_{\rm e,\,inj}(r) \propto L_{\rm tot,\,FC}(r) \simeq L_{\gamma}(r) +
L_{\rm opt}(r) \sim L_{\gamma}(r) \, .
\end{equation}
while the optical luminosity is
\begin{equation}
L_{\rm opt}(r) \simeq \frac{u_{\rm B}'(r)}{u_{\rm ext}'(r)} \,
L_{\gamma}(r) \propto \frac{u_{\rm B}'(r)}{u_{\rm ext}'(r)} \, P_{\rm
e,\,inj}(r) \, ,
\end{equation}
where we assumed $\delta = \Gamma_{\rm j}$ being independent on the
position $r$ along the jet. Hence, it is clear that while a maximum of
$L_{\gamma}(r)$ is determined solely by the injection rate $P_{\rm
e,\,inj}(r)$, a maximum of $L_{\rm opt}(r)$ may in general be quite different,
depending on particular radial profiles of $P_{\rm e,\,inj}(r)$ and of the
ratio $[u'_{\rm B}/u'_{\rm ext}]\,\left(r\right)$.

As a specific illustrative example, let us assume that the dissipation
region propagating down the jet injects non-thermal energy into radiating
particles at the rate being a broad Gaussian function of distance $r$ with
a maximum at $r_0$ and a width of $r_0/\sqrt{2}$,
\begin{equation}
P_{\rm e,\,inj} \propto \exp{[-(r-r_0)^2/r_0^2]} \, ,
\end{equation}
and that magnetic field and external photon field energy densities scale
with $r$ as power laws with indices $\beta_{\rm B}$ and $\beta_{\rm ext}$
\begin{equation}
u'_{\rm B}(r) \propto r^{-\beta_{\rm B}} \quad {\rm and}  \quad u'_{\rm ext}(r)
\propto r^{-\beta_{\rm ext}} \, .
\end{equation}
Then one can find that $L_{\gamma}$ has a maximum at $r=r_0$, as expected,
whereas $L_{\rm opt}$ attains a maximum at
\begin{equation}
r_{\rm cr} = \frac{r_0}{2} \times (1 + \sqrt{1+2\,(\beta_{\rm ext}-\beta_B)}) \, ,
\end{equation}
which is larger than $r_0$ as long as $\beta_{\rm ext} > \beta_{\rm B}$ and thus results in the
optical flare lagging the $\gamma$-ray flare. This is due to the
fact that with the magnetic energy density decreasing less rapidly in the
jet comoving frame than the external radiation energy density, the drop in
the injection rate $P_{\rm e,\,inj}(r)$ between $r_0$ and $r_{\rm cr}$ is
compensated by the \emph{increase} in the ratio $[u'_{\rm B}/u'_{\rm
ext}]\,\left(r\right)$.

Let us further consider the particular values of $\beta_{\rm B}=2$ and $\beta_{\rm
ext} =4$ discussed in Section~\ref{section:modeling}. With such, assuming
again the electron injection rate being a broad Gaussian function of the
distance $r$ along the jet as in the example above, the observed time lag
between the optical and $\gamma$-ray flares $\Delta t_{obs}$ can be
evaluated as roughly
\begin{equation}
\Delta t_{obs} \simeq \frac{0.6 \, r_0}{c \, \Gamma_{\rm j}^2} \simeq 3 \times
\left(\frac{\Gamma_{\rm j}}{15}\right)^{-2} \, \left(\frac{r_0}{1\,{\rm
pc}}\right) \, {\rm days} \, .
\end{equation}
It is encouraging that a 10-day lag is expected for
$\Gamma_{\rm j} \simeq 15$ and $r_0 \simeq 3$\,pc, which are the bulk
Lorentz factor and the location of the dissipation region comparable to
that inferred from our ERC-IR modeling.\\

%


\end{document}